%% file: ms.tex
\documentclass[reprint,aps,prd,twocolumn ,groupedaddress,nobibnotes]{revtex4-1}
\usepackage[T1]{fontenc}
\usepackage[latin9]{inputenc}
\usepackage{booktabs}
\usepackage{amsmath}
\usepackage{amssymb}
\usepackage{graphicx}
\pdfoutput=1

\newcommand{\nubar}{$\bar\nu_\mu \ $}

\newcommand{\nubarandnu}{$\bar\nu_\mu$ and $\nu_\mu \ $}
\newcommand{\nuandnubar}{$\nu_\mu$ and $\bar\nu_\mu \ $}

\makeatletter

\providecommand{\tabularnewline}{\\}

\usepackage{lineno}

\usepackage{soul}
\begin{document}

\title{Measurement of $\overline{\nu}_{\mu}$ and $\nu_{\mu}$ charged current inclusive cross sections
and their ratio with the T2K off-axis near detector}


\input{./latex-author.tex}




\date{\today}
\begin{abstract}
We report a measurement of cross section $\sigma(\nu_{\mu}+{\rm nucleus}\rightarrow\mu^{-}+X)$
and the first measurements of the cross section $\sigma(\overline{\nu}_{\mu}+{\rm nucleus}\rightarrow\mu^{+}+X)$
and their ratio 
 $R(\frac{\sigma(\bar \nu)}{\sigma(\nu)})$
at 
(anti-)neutrino
energies below 1.5 GeV. We determine the single momentum bin cross section
measurements, averaged over the T2K $\overline{\nu}/\nu$-flux, for the
detector target material (mainly Carbon, Oxygen, Hydrogen and Copper)
with phase space restricted laboratory frame kinematics of 
$\theta_{\mu}<32^\circ$
and $p_{\mu}>$500 MeV/c. 
The results are
$\sigma(\overline{\nu})=\left(
0.900\pm0.029{\rm (stat.)}\pm0.088{\rm (syst.)}\right)\times10^{-39}$
and 
$\sigma(\nu)=\left( 2.41\ \pm0.022{\rm{(stat.)}}\pm0.231{\rm (syst.)}\ \right)\times10^{-39}$
in units of cm$^{2}$/nucleon 
and 
$R\left(\frac{\sigma(\overline{\nu})}{\sigma(\nu)}\right)=
0.373\pm0.012{\rm (stat.)}\pm0.015{\rm (syst.)}$.

\begin{description}
\item [{PACS~numbers}] \textsf{13.15.+g, 14.60.Pq, 14.60.Lm, 25.30.Pt, 29.40.Mc}.{\small \par}
\end{description}
\end{abstract}


\maketitle
\input{./Sections/Introduction-b.tex}

\input{./Sections/DetectorAndBeam-a.tex}

\input{./Sections/AnalysisSamples.tex}

\input{./Sections/EventSelection-a.tex}

\input{./Sections/AnalysisMethods-2.tex}
\input{./Sections/PropagationOfSystematics-b.tex} 
\input{./Sections/Results-a-Erez_Jan24.tex}
\input{./Sections/Conclusions.tex}

\medskip

\begin{acknowledgments}
We thank the J-PARC staff for superb accelerator performance. We thank the 
CERN NA61/SHINE Collaboration for providing valuable particle production data.
We acknowledge the support of MEXT, Japan; 
NSERC (Grant No. SAPPJ-2014-00031), NRC and CFI, Canada;
CEA and CNRS/IN2P3, France;
DFG, Germany; 
INFN, Italy;
National Science Centre (NCN) and Ministry of Science and Higher Education, Poland;
RSF, RFBR, and MES, Russia; 
MINECO and ERDF funds, Spain;
SNSF and SERI, Switzerland;
STFC, UK; and 
DOE, USA.
We also thank CERN for the UA1/NOMAD magnet, 
DESY for the HERA-B magnet mover system, 
NII for SINET4, 
the WestGrid and SciNet consortia in Compute Canada, 
and GridPP in the United Kingdom.
In addition, participation of individual researchers and institutions has been further 
supported by funds from ERC (FP7), H2020 Grant No. RISE-GA644294-JENNIFER, EU; 
JSPS, Japan; 
Royal Society, UK; 
the Alfred P. Sloan Foundation and the DOE Early Career program, USA.
\end{acknowledgments}

\end{document}

%% file: latex-author.tex

\newcommand{\INSTEE}{\affiliation{University of Bern, Albert Einstein Center for Fundamental Physics, Laboratory for High Energy Physics (LHEP), Bern, Switzerland}}
\newcommand{\INSTFE}{\affiliation{Boston University, Department of Physics, Boston, Massachusetts, U.S.A.}}
\newcommand{\INSTD}{\affiliation{University of British Columbia, Department of Physics and Astronomy, Vancouver, British Columbia, Canada}}
\newcommand{\INSTGA}{\affiliation{University of California, Irvine, Department of Physics and Astronomy, Irvine, California, U.S.A.}}
\newcommand{\INSTI}{\affiliation{IRFU, CEA Saclay, Gif-sur-Yvette, France}}
\newcommand{\INSTGB}{\affiliation{University of Colorado at Boulder, Department of Physics, Boulder, Colorado, U.S.A.}}
\newcommand{\INSTFG}{\affiliation{Colorado State University, Department of Physics, Fort Collins, Colorado, U.S.A.}}
\newcommand{\INSTFH}{\affiliation{Duke University, Department of Physics, Durham, North Carolina, U.S.A.}}
\newcommand{\INSTBA}{\affiliation{Ecole Polytechnique, IN2P3-CNRS, Laboratoire Leprince-Ringuet, Palaiseau, France }}
\newcommand{\INSTEG}{\affiliation{University of Geneva, Section de Physique, DPNC, Geneva, Switzerland}}
\newcommand{\INSTDG}{\affiliation{H. Niewodniczanski Institute of Nuclear Physics PAN, Cracow, Poland}}
\newcommand{\INSTCB}{\affiliation{High Energy Accelerator Research Organization (KEK), Tsukuba, Ibaraki, Japan}}
\newcommand{\INSTED}{\affiliation{Institut de Fisica d'Altes Energies (IFAE), The Barcelona Institute of Science and Technology, Campus UAB, Bellaterra (Barcelona) Spain}}
\newcommand{\INSTEC}{\affiliation{IFIC (CSIC \& University of Valencia), Valencia, Spain}}
\newcommand{\INSTEI}{\affiliation{Imperial College London, Department of Physics, London, United Kingdom}}
\newcommand{\INSTGF}{\affiliation{INFN Sezione di Bari and Universit\`a e Politecnico di Bari, Dipartimento Interuniversitario di Fisica, Bari, Italy}}
\newcommand{\INSTBE}{\affiliation{INFN Sezione di Napoli and Universit\`a di Napoli, Dipartimento di Fisica, Napoli, Italy}}
\newcommand{\INSTBF}{\affiliation{INFN Sezione di Padova and Universit\`a di Padova, Dipartimento di Fisica, Padova, Italy}}
\newcommand{\INSTBD}{\affiliation{INFN Sezione di Roma and Universit\`a di Roma ``La Sapienza'', Roma, Italy}}
\newcommand{\INSTEB}{\affiliation{Institute for Nuclear Research of the Russian Academy of Sciences, Moscow, Russia}}
\newcommand{\INSTHA}{\affiliation{Kavli Institute for the Physics and Mathematics of the Universe (WPI), The University of Tokyo Institutes for Advanced Study, University of Tokyo, Kashiwa, Chiba, Japan}}
\newcommand{\INSTCC}{\affiliation{Kobe University, Kobe, Japan}}
\newcommand{\INSTCD}{\affiliation{Kyoto University, Department of Physics, Kyoto, Japan}}
\newcommand{\INSTEJ}{\affiliation{Lancaster University, Physics Department, Lancaster, United Kingdom}}
\newcommand{\INSTFC}{\affiliation{University of Liverpool, Department of Physics, Liverpool, United Kingdom}}
\newcommand{\INSTFI}{\affiliation{Louisiana State University, Department of Physics and Astronomy, Baton Rouge, Louisiana, U.S.A.}}
\newcommand{\INSTHB}{\affiliation{Michigan State University, Department of Physics and Astronomy,  East Lansing, Michigan, U.S.A.}}
\newcommand{\INSTCE}{\affiliation{Miyagi University of Education, Department of Physics, Sendai, Japan}}
\newcommand{\INSTDF}{\affiliation{National Centre for Nuclear Research, Warsaw, Poland}}
\newcommand{\INSTFJ}{\affiliation{State University of New York at Stony Brook, Department of Physics and Astronomy, Stony Brook, New York, U.S.A.}}
\newcommand{\INSTGJ}{\affiliation{Okayama University, Department of Physics, Okayama, Japan}}
\newcommand{\INSTCF}{\affiliation{Osaka City University, Department of Physics, Osaka, Japan}}
\newcommand{\INSTGG}{\affiliation{Oxford University, Department of Physics, Oxford, United Kingdom}}
\newcommand{\INSTBB}{\affiliation{UPMC, Universit\'e Paris Diderot, CNRS/IN2P3, Laboratoire de Physique Nucl\'eaire et de Hautes Energies (LPNHE), Paris, France}}
\newcommand{\INSTGC}{\affiliation{University of Pittsburgh, Department of Physics and Astronomy, Pittsburgh, Pennsylvania, U.S.A.}}
\newcommand{\INSTFA}{\affiliation{Queen Mary University of London, School of Physics and Astronomy, London, United Kingdom}}
\newcommand{\INSTE}{\affiliation{University of Regina, Department of Physics, Regina, Saskatchewan, Canada}}
\newcommand{\INSTGD}{\affiliation{University of Rochester, Department of Physics and Astronomy, Rochester, New York, U.S.A.}}
\newcommand{\INSTHC}{\affiliation{Royal Holloway University of London, Department of Physics, Egham, Surrey, United Kingdom}}
\newcommand{\INSTBC}{\affiliation{RWTH Aachen University, III. Physikalisches Institut, Aachen, Germany}}
\newcommand{\INSTFB}{\affiliation{University of Sheffield, Department of Physics and Astronomy, Sheffield, United Kingdom}}
\newcommand{\INSTDI}{\affiliation{University of Silesia, Institute of Physics, Katowice, Poland}}
\newcommand{\INSTEH}{\affiliation{STFC, Rutherford Appleton Laboratory, Harwell Oxford,  and  Daresbury Laboratory, Warrington, United Kingdom}}
\newcommand{\INSTCH}{\affiliation{University of Tokyo, Department of Physics, Tokyo, Japan}}
\newcommand{\INSTBJ}{\affiliation{University of Tokyo, Institute for Cosmic Ray Research, Kamioka Observatory, Kamioka, Japan}}
\newcommand{\INSTCG}{\affiliation{University of Tokyo, Institute for Cosmic Ray Research, Research Center for Cosmic Neutrinos, Kashiwa, Japan}}
\newcommand{\INSTGI}{\affiliation{Tokyo Metropolitan University, Department of Physics, Tokyo, Japan}}
\newcommand{\INSTF}{\affiliation{University of Toronto, Department of Physics, Toronto, Ontario, Canada}}
\newcommand{\INSTB}{\affiliation{TRIUMF, Vancouver, British Columbia, Canada}}
\newcommand{\INSTG}{\affiliation{University of Victoria, Department of Physics and Astronomy, Victoria, British Columbia, Canada}}
\newcommand{\INSTDJ}{\affiliation{University of Warsaw, Faculty of Physics, Warsaw, Poland}}
\newcommand{\INSTDH}{\affiliation{Warsaw University of Technology, Institute of Radioelectronics, Warsaw, Poland}}
\newcommand{\INSTFD}{\affiliation{University of Warwick, Department of Physics, Coventry, United Kingdom}}
\newcommand{\INSTGH}{\affiliation{University of Winnipeg, Department of Physics, Winnipeg, Manitoba, Canada}}
\newcommand{\INSTEA}{\affiliation{Wroclaw University, Faculty of Physics and Astronomy, Wroclaw, Poland}}
\newcommand{\INSTHE}{\affiliation{Yokohama National University, Faculty of Engineering, Yokohama, Japan}}
\newcommand{\INSTH}{\affiliation{York University, Department of Physics and Astronomy, Toronto, Ontario, Canada}}

\INSTEE
\INSTFE
\INSTD
\INSTGA
\INSTI
\INSTGB
\INSTFG
\INSTFH
\INSTBA
\INSTEG
\INSTDG
\INSTCB
\INSTED
\INSTEC
\INSTEI
\INSTGF
\INSTBE
\INSTBF
\INSTBD
\INSTEB
\INSTHA
\INSTCC
\INSTCD
\INSTEJ
\INSTFC
\INSTFI
\INSTHB
\INSTCE
\INSTDF
\INSTFJ
\INSTGJ
\INSTCF
\INSTGG
\INSTBB
\INSTGC
\INSTFA
\INSTE
\INSTGD
\INSTHC
\INSTBC
\INSTFB
\INSTDI
\INSTEH
\INSTCH
\INSTBJ
\INSTCG
\INSTGI
\INSTF
\INSTB
\INSTG
\INSTDJ
\INSTDH
\INSTFD
\INSTGH
\INSTEA
\INSTHE
\INSTH

\author{K.\,Abe}\INSTBJ
\author{J.\,Amey}\INSTEI
\author{C.\,Andreopoulos}\INSTEH\INSTFC
\author{M.\,Antonova}\INSTEB
\author{S.\,Aoki}\INSTCC
\author{A.\,Ariga}\INSTEE
\author{Y.\,Ashida}\INSTCD
\author{S.\,Ban}\INSTCD
\author{M.\,Barbi}\INSTE
\author{G.J.\,Barker}\INSTFD
% ...

\collaboration{The T2K Collaboration}\noaffiliation

%% file: Sections/Introduction-b.tex
\section{Introduction}

Since the 1998 discovery \citep{fukuda} of neutrino oscillations,
there have been major advances in neutrino disappearance
and appearance  oscillation measurements
and all the fundamental neutrino mixing parameters \citep{PDG-2016} have been determined
except for the mass hierarchy and the charge-parity (CP)
phase $\delta_{CP}$. 
Evidence of $\delta_{CP}\ne 0, \pi$  leads to the non-conservation or violation
of the charge-parity symmetry (CPV).
This is tested by measuring the neutrino $\nu_{\mu}\rightarrow\nu_{e}$ and antineutrino
$\overline{\nu}_{\mu}\rightarrow\overline{\nu}_{e}$ appearance oscillation
event rates to determine if the neutrino and antineutrino oscillation appearance probabilities, 
$P\left(\nu_{\mu}\rightarrow\nu_{e}\right)$
and
$\bar{P}\left(\overline{\nu}_{\mu}\rightarrow\overline{\nu}_{e}\right)$
are equal in vacuum \citep{msw}
at the same ratio of  
the oscillation distance $L$
over
the neutrino energy $E$ or $\frac{L}{E}$.
Major 
long-baseline neutrino experiments \citep{t2k+nova} have been built
and future projects \citep{dune+hyperk} are proposed to determine
these probabilities  using separate $\nu_{\mu}$ and $\overline{\nu}_{\mu}$ beams
that cross near and far detectors. 
The probabilities are obtained from near detector measurements of the 
$\nu_{\mu}+N$ and $\overline{\nu}_{\mu}+N$ charged current (CC) interactions and cross sections,
where $N$ is the target nucleon, 
and far detector measurements of $\nu_{e}+N$ and $\overline{\nu}_{e}+N$ CC interactions. 

In this paper, the T2K Collaboration, using the off-axis near detector (ND280), presents
a measurement at a peak energy 
$\sim$0.6 GeV of the 
charged current inclusive (CCINC)
$\nu_{\mu}+N$ cross section and first CCINC
measurements of the $\overline{\nu}_{\mu}+N$ cross section and
their ratio of the  $\overline{\nu}_{\mu}+N$ over the $\nu_{\mu}+N$
CCINC cross section.
These $\nu_{\mu}$  and $\bar\nu_{\mu}$ measurements 
are important to understand their
impact on future CPV measurements and to 
test neutrino cross section models.

T2K has published flux averaged neutrino-mode  measurements  of CCINC \citep{t2k-carbon-ccinc}
and charged current quasi-elastic like (CCQE) \citep{t2k-cceq}
cross sections per nucleon of  
 $(6.91\pm0.13{\rm (stat.)}\pm0.84{\rm (syst.)}) \times 10^{-39}$cm$^2$ 
and $(4.15\pm0.6) \times 10^{-39}$cm$^2$, 
respectively. These measurements were performed using the
Fine-Grain Detector (FGD) which has different detector systematics
compared to the measurements presented in this paper.
There are no published CCINC $\bar\nu_{\mu}$
measurements at
energies below 1.5 GeV, however
the MINVERVA Collaboration recent published \citep{minerva+nu+antinu}
CCINC results above 2 GeV
and
the MiniBooNE Collaboration has published \citep{miniboone} CCQE
measurements in both 
\nubarandnu
modes which require
larger axial mass values compared to other experiments to fit their observed data. There are several
multinucleon models (2 particle 2 hole, or 2p2h)
 \citep{martini-1,nieves-1,martini-2} proposed to explain large cross sections.
In addition, in some models it has been predicted \citep{martini-1} that the difference between the 
\nuandnubar 
cross sections is expected to increase when 
2p2h effects \citep{2p2h-refs}
are included. The measurements of the ratio, sum, and difference of these cross sections,  which
have very different systematic errors, will be presented.

Following this introduction, the paper is organized as follows. We
begin with a description of the ND280 off-axis detector and the neutrino
beam in Section II. Then the Monte Carlo (MC) simulation is presented in
Section III, followed by the event selection given in Section IV.
The analysis methods and systematic error evaluations are presented
in Sections V and VI and we finally conclude with the Results and
Conclusions in Sections VII and VIII.

%% file: Sections/DetectorAndBeam-a.tex
\section{Beam and Detector}

The T2K experiment \citep{t2k-expt} is composed of a neutrino beamline
and a near detector at the J-PARC laboratory in Tokai, Japan, and
the far detector Super-Kamiokande (SK) situated \mbox{295 km} away in the Kamioka mine. 
The J-PARC accelerator complex produces a 30 GeV energy proton beam with spills
every 2.48 s that contain eight beam bunches which are 580 ns apart.
At this spill and repetition rate, 
a beam power of 430 kW
produces $2.25\times 10^{14}$ protons on target (PoT) per spill
corresponding to $\approx 0.8\times 10^{19}$ PoT integrated 
per day of data taking.

The proton beam
strikes a graphite target to produce pions and kaons that are focused
by three magnetic horns into a 96 m long decay pipe. The polarity of the magnetic horns
can be changed to Forward Horn Current (FHC) or Reverse
Horn Current (RHC) to select either positive or negative pions and
kaons to produce a predominantly $\nu_\mu$ or an $\bar\nu_\mu$ beam.
%
%
The resulting main neutrino beam axis is parallel to the proton beam direction. 
SK lies 2.5$^\circ$ off-axis with respect to the
main neutrino beam direction and this arrangement produces at SK both
the $\nu_\mu$ and $\bar\nu_\mu$ energies that peak at $\sim$0.6 GeV.
This $\nu_\mu$($\overline\nu_\mu$) peak energy with a 295 km baseline distance,
produces an $\frac{L}{E}$ value that maximizes the $\nu_{e}$($\overline{\nu}_{e}$)
appearance rate and has a $\nu_{\mu}$($\overline{\nu}_{\mu}$) disappearance
that minimizes the $\nu_\mu$($\overline\nu_\mu$) rates at SK.

\begin{figure}
\includegraphics[scale=1.0]{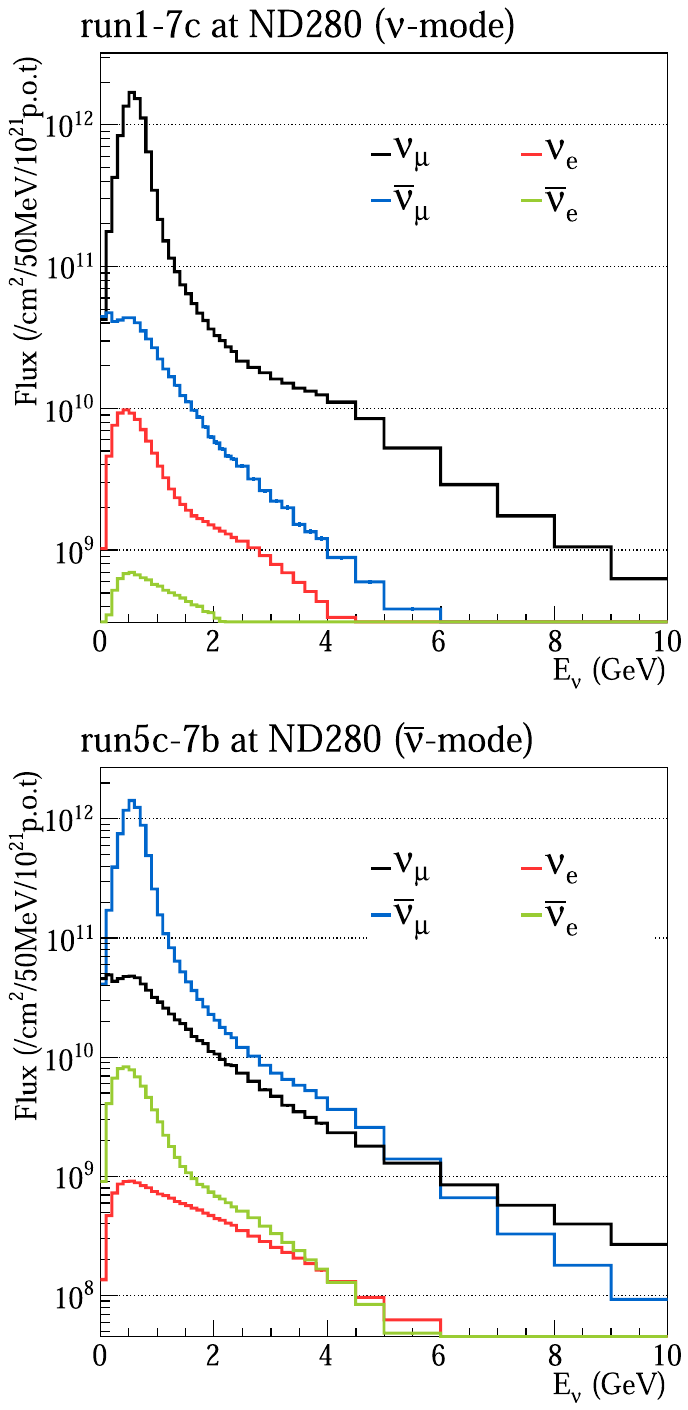} 
\caption{\label{fig:The-neutrino-FHC}The predominately neutrino FHC beam (Top) and 
predominately antineutrino
RHC beam (Bottom) flux per PoT as a function of energy
at the ND280 detector. The rates are separated by neutrino/antineutrino
muon and electron type flavors. The peak values for the neutrino and
the antineutrino flux rates are $1.7\times10^{12}$ and $1.4\times10^{12}$
/cm$^{2}$/50MeV/10$^{21}$ PoT, respectively.}
\end{figure}

The ND280 \nuandnubar fluxes were determined by simulation
of the T2K neutrino beamline \citep{t2k-flux} using 
FLUKA2011 \citep{fluka}, 
GEANT \citep{geant},
and GCALOR \citep{gcalor} 
software packages. The simulated hadronic yields have been re-weighted
using the 
NA61/SHINE \citep{shine} thin-target data,
which has reduced the flux uncertainties to less than 10\% around the flux peak.
Detailed
descriptions of the ND280 flux uncertainties have been published in
previous ND280 analyses \citep{tpc-analysis-paper}. 
The typical fractional covariance
error of the T2K \nuandnubar fluxes are $\sim$10\%
and the $\nu_\mu$-$\bar\nu_\mu$ correlated flux errors are $\sim$6\%.
The \nuandnubar flux rates
per cm$^{2}$/50 MeV/$10^{21}$ PoT are plotted in Fig.\ref{fig:The-neutrino-FHC}
with superimposed neutral lepton flavors, $\nu_\mu$,  $\nu_e$, $\bar\nu_\mu$ and $\bar\nu_e$.

The near detector complex, located 280m downstream of the target, consists
of an on-axis
detector (INGRID) and the ND280 off-axis detector. 
ND280 is positioned 
inline between the neutrino beam target and SK.  
The ND280 detector consists of sub-detectors inside
the refurbished UA1/NOMAD magnet that operates at a 0.2 T magnetic
field whose direction is horizontal and perpendicular to the neutrino beam.
The ND280 sub-detectors include $\pi^0$ detector \citep{p0d-1} (P$\emptyset$D),
three tracking time projection chambers \citep{tpc} (TPC1,2,3), two
fine-grained detectors (FGD1,2) interleaved with TPC1,2,3, and an
electromagnetic calorimeter (ECAL), that encloses the P$\emptyset$D, TPC1-3
and FGD1-2 sub-detectors. 

The measurements in this paper used the P$\emptyset$D and the TPC
tracking sub-detectors in the ND280 detector complex. In our description,
the +Z direction is parallel to the neutrino beam direction, and the +Y direction
is vertically upwards. Previous descriptions of analyses using the P$\emptyset$D have been published \citep{pod-nue}.
We describe additional details relevant for the analysis presented in this paper.

\begin{figure}
\includegraphics[scale=0.4]{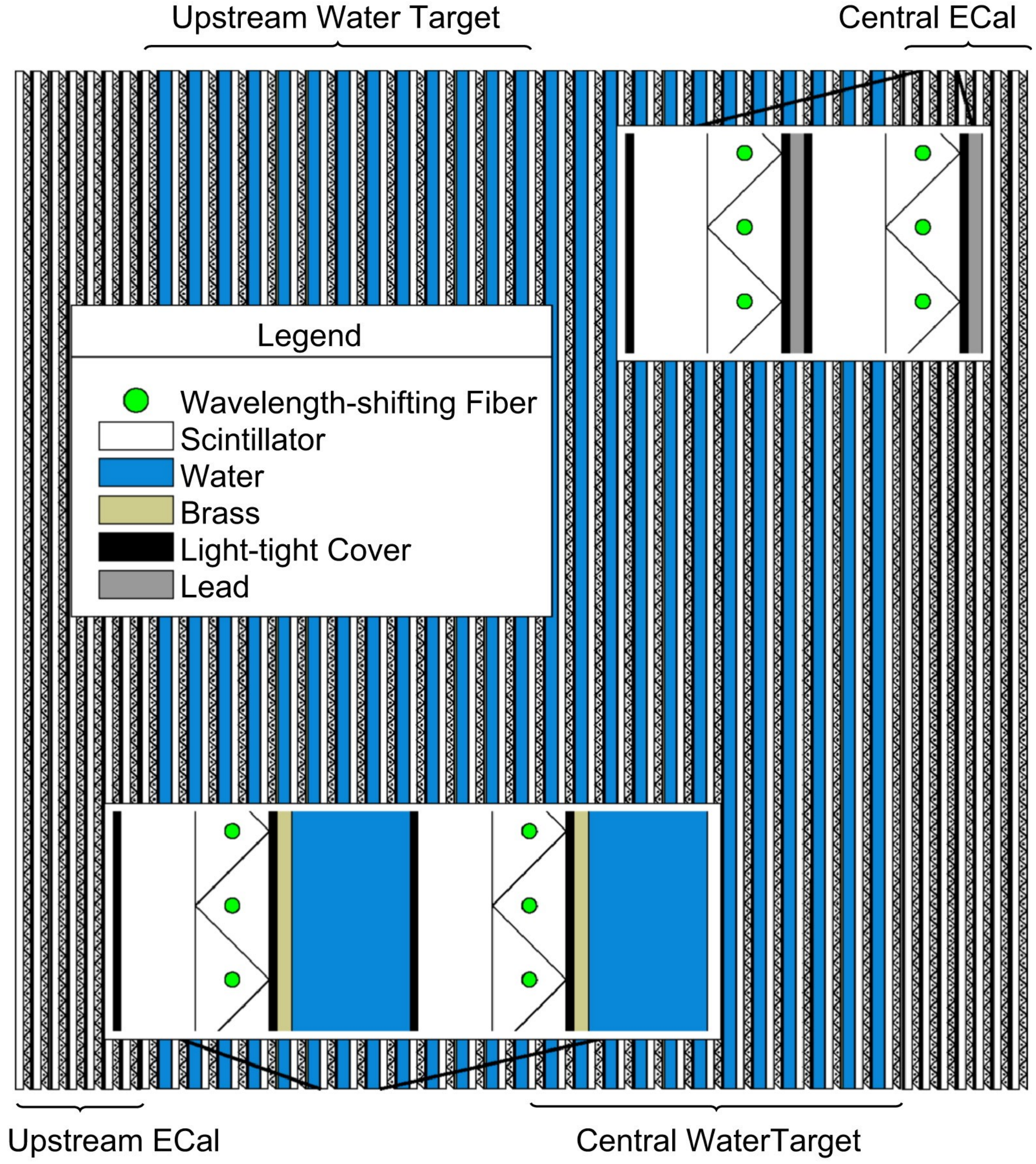}

\caption{\label{fig:Side-view-schematic}Side view schematic diagram of the
P$\emptyset$D detector. The white, zig-zag, and blue strip regions represent
the vertical scintillator bars, the horizontal scintillator bars,
and the water bag regions, respectively. The vertical and horizontal
bars represent a X-Y module or P$\emptyset$Dule. The first and last groups of
seven P$\emptyset$Dules form the upstream and the central ECAL ``super'' modules
and the middle 26 P$\emptyset$Dules interleaved with the water bags are the
water target region.}
\end{figure}

The P$\emptyset$D is shown in Fig.\ref{fig:Side-view-schematic}.
This detector contains 40 scintillator module planes called P$\emptyset$Dules.
Each P$\emptyset$Dule has 134 horizontal and 126 vertical triangular scintillator bars. 
A wavelength
shifting fiber centered in each bar is readout on one end by a silicon photomultiplier.
The P$\emptyset$D dimensions are $2298\times2468\times2350$ mm$^{3}$\textemdash XYZ\textemdash with
a total mass of $\sim$1900 kg of water and 3570 kg of other materials (mainly
scintillator with thin layers of high density polyethylene plastic and brass
sheet). The target material mass is given in fractional amounts in
Table \ref{tab:Chemical-composition-of}. These P$\emptyset$Dules are formed
into 3 major sections. The water target region, is
the primary target in this analysis which has 26 P$\emptyset$Dules interleaved with
bags of water 2.8 cm thick and 1.3 mm brass sheets. 
The water bags are drainable to allow water target subtraction
measurements. The two other regions (called upstream and central ECALs)
are the upstream and downstream sections that each contain 7 P$\emptyset$Dules
and steel sheets clad with lead (4.9 radiation lengths). 

\begin{table}
\caption{\label{tab:Chemical-composition-of}Chemical element composition of
P$\emptyset$D water target region by fraction of mass. }

\begin{tabular}{lcc}
\hline 
\hline
Element & Symbol & Fraction\tabularnewline 
\hline 
Hydrogen & H & 8.0\%\tabularnewline 
Carbon & C & 45.0\%\tabularnewline 
Oxygen & O & 29.9\%\tabularnewline 
Copper & Cu & 14.3\%\tabularnewline 
Chlorine & Cl & 1.1\%\tabularnewline 
Titanium & Ti & 0.1\%\tabularnewline 
Zinc & Zn & 1.6\%\tabularnewline
\hline
\hline 
\end{tabular}

\end{table}

The TPC1,2,3 detectors are three modules whose dimensions are each
$1808\times2230\times852$ mm$^{3}$\textemdash XYZ\textemdash where each module
contains a centered high voltage (Z-Y)
cathode plane that splits the chamber into two sections where the charged particle track ionizations drift
in the $\pm X$ directions. These are measured by 70~mm$^{2}$ micromegas
pads in the Z-Y plane. The fully contained ionized track path
lengths are 72 cm. A charged track will be measured with $\sim$0.7
mm resolution for drift distances >10 cm.
The typical
TPC momentum resolution is $\delta\left(p_{\bot}\right)/p_{\bot}=0.08p_{\bot}$ (GeV/c). 
Analyses which use the TPC have been
described in previous ND280 publications \citep{tpc-analysis-paper}.


%% file: Sections/AnalysisSamples.tex
\section{Analysis Samples}

The studies reported here includes data logged with 
the FHC $\nu$ beam runs (
October 2012 to February 2013
)
and the  RHC $\overline{\nu}$ beam runs
(May 2014 to June 2014).

\subsection{Data samples and detector configuration}
The total PoT exposure where all detector data quality checks were passed 
for the FHC runs was 16.24$\times10^{19}$ 
and the corresponding total PoT exposure for the RHC runs was 4.30$\times 10^{19}$.
These integrated rates corresponds to roughly
$0.28\times10^{12}$ neutrinos and $0.06\times10^{12}$ antineutrinos
per cm$^{2}$ per 50 MeV at 0.6 GeV. 
The data samples in this paper used the available neutrino and antineutrino beam 
data taken when the P$\emptyset$D target bags were filled with water.  

\subsection{Monte Carlo simulation}

The analysis used simulated MC
samples with different beam and detector configurations for each data taking period. 
The simulations include the following: 
\begin{enumerate}
\item
Secondary pions and kaons are produced in the graphite target and propagated through 
the magnetic horns into a helium filled pipe where 
they decay.  Secondary neutrinos and antineutrinos are created
and their fluxes and energy spectra 
are extrapolated to the near and far detectors.
\item
The neutrino and antineutrino interactions in the ND280 sub-detectors
were determined by the NEUT \citep{NEUT-1} MC generator 
that was used to calculate the interaction cross sections and 
the final state particle kinematics.
\item
The detector simulation uses GEANT to propagate the final state particles 
through the ND280 sub-detectors.
\end{enumerate}

%% file: Sections/EventSelection-a.tex
\section{Event and Kinematic Selection}

\subsection{Event selection}


The analysis selection uses reconstructed objects from both the P$\emptyset$D
and TPC. Both sub-detectors use independent reconstruction algorithms
to generate objects from the raw data. The P$\emptyset$D uses a 3D tracking
algorithm to form tracks from individual hits in the scintillator
bars. The TPC reconstruction uses a track in the Y-Z plane (non-drift plane)
as a seed to search for hits in the downstream FGD to form a track object.
\begin{figure*}
\includegraphics[scale=0.29]{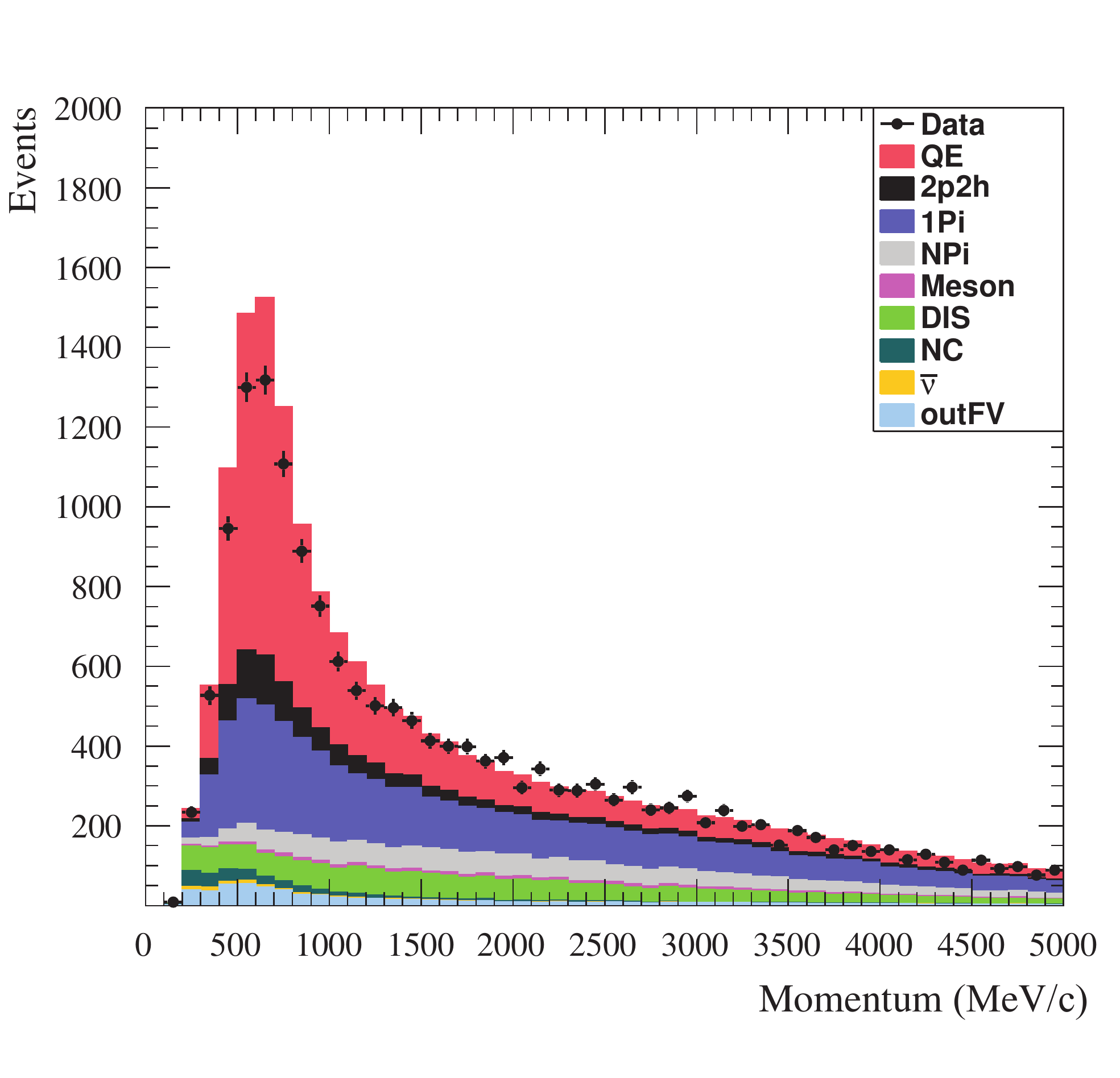} \includegraphics[scale=0.29]{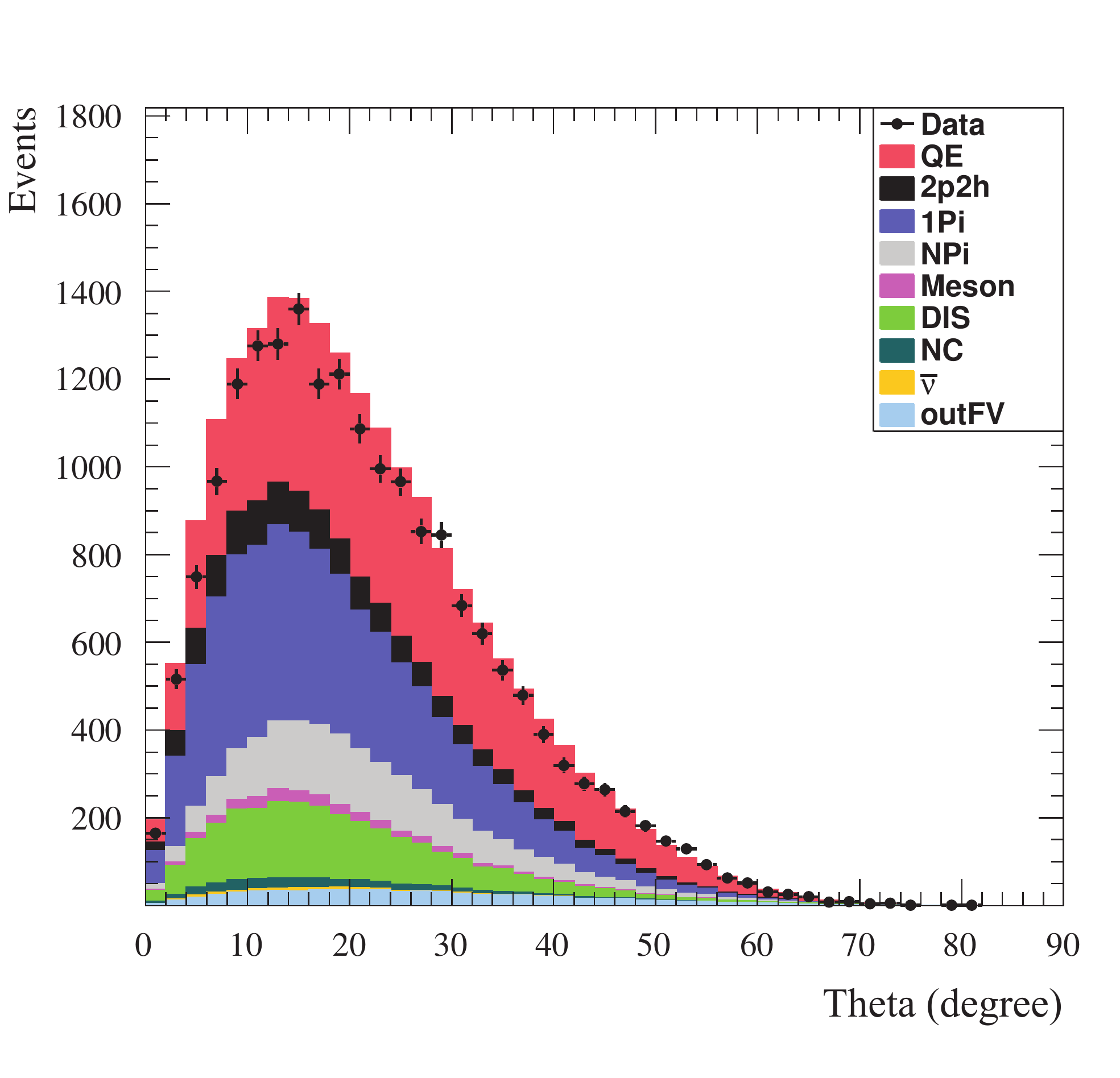} \includegraphics[scale=0.29]{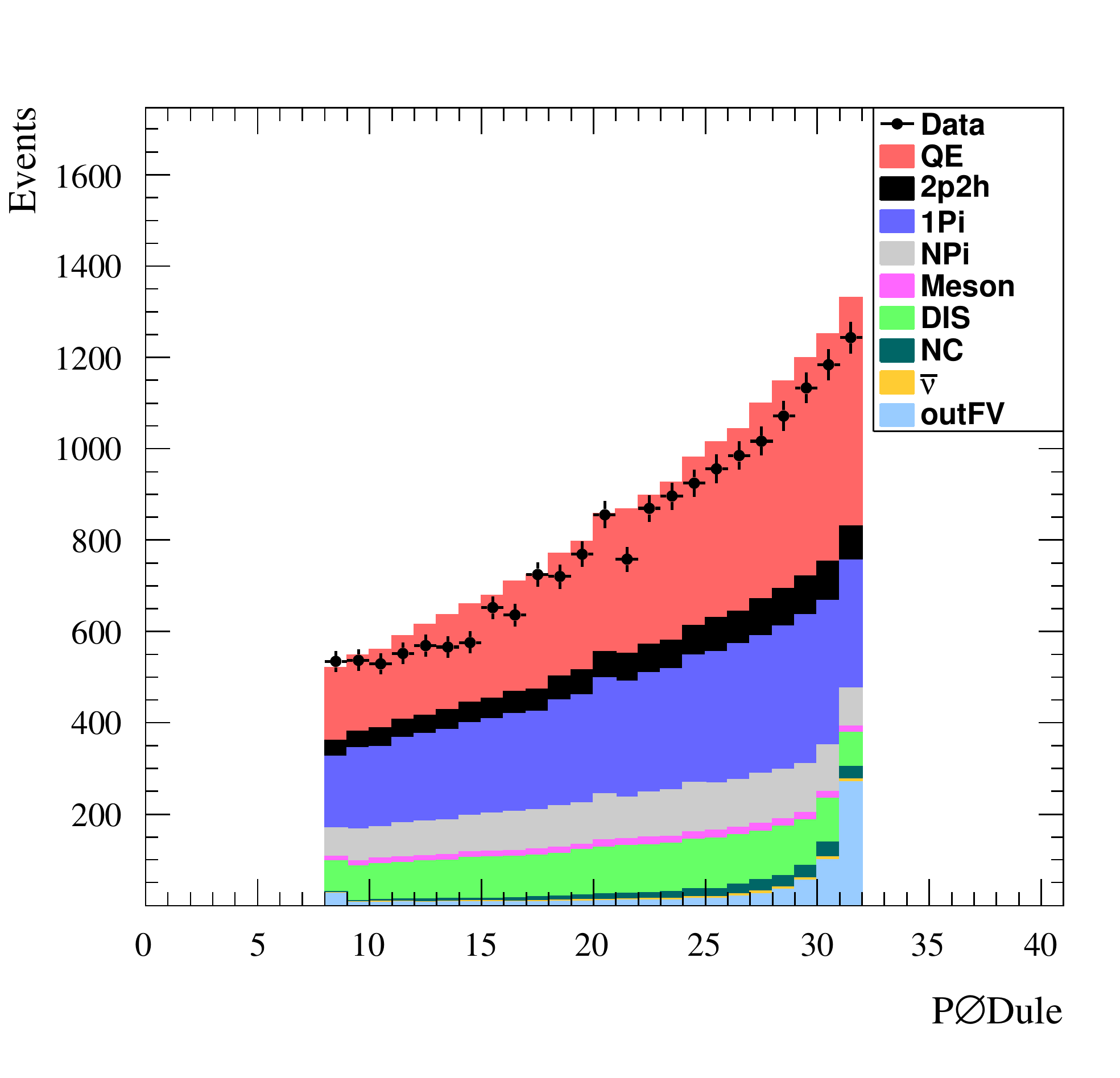}

\caption{\label{fig:FHC-Neutrino-beam}FHC beam CCINC $\nu_\mu$ event
candidate distributions of the $\mu^{-}$ momentum in MeV/c
(Left), the muon $\theta_{\mu}$ angle in degrees (Middle), and interaction
vertex position by P$\emptyset$Dule (Right). Note backgrounds in the
CCINC sample are the NC (dark green), $\bar\nu_\mu$ induced events (yellow)
and the out of fiducial volume events (light blue).
There are negligible $\bar\nu_\mu$ backgrounds (yellow)
in the FHC sample.}
\end{figure*}

\begin{figure*}
\includegraphics[scale=0.29]{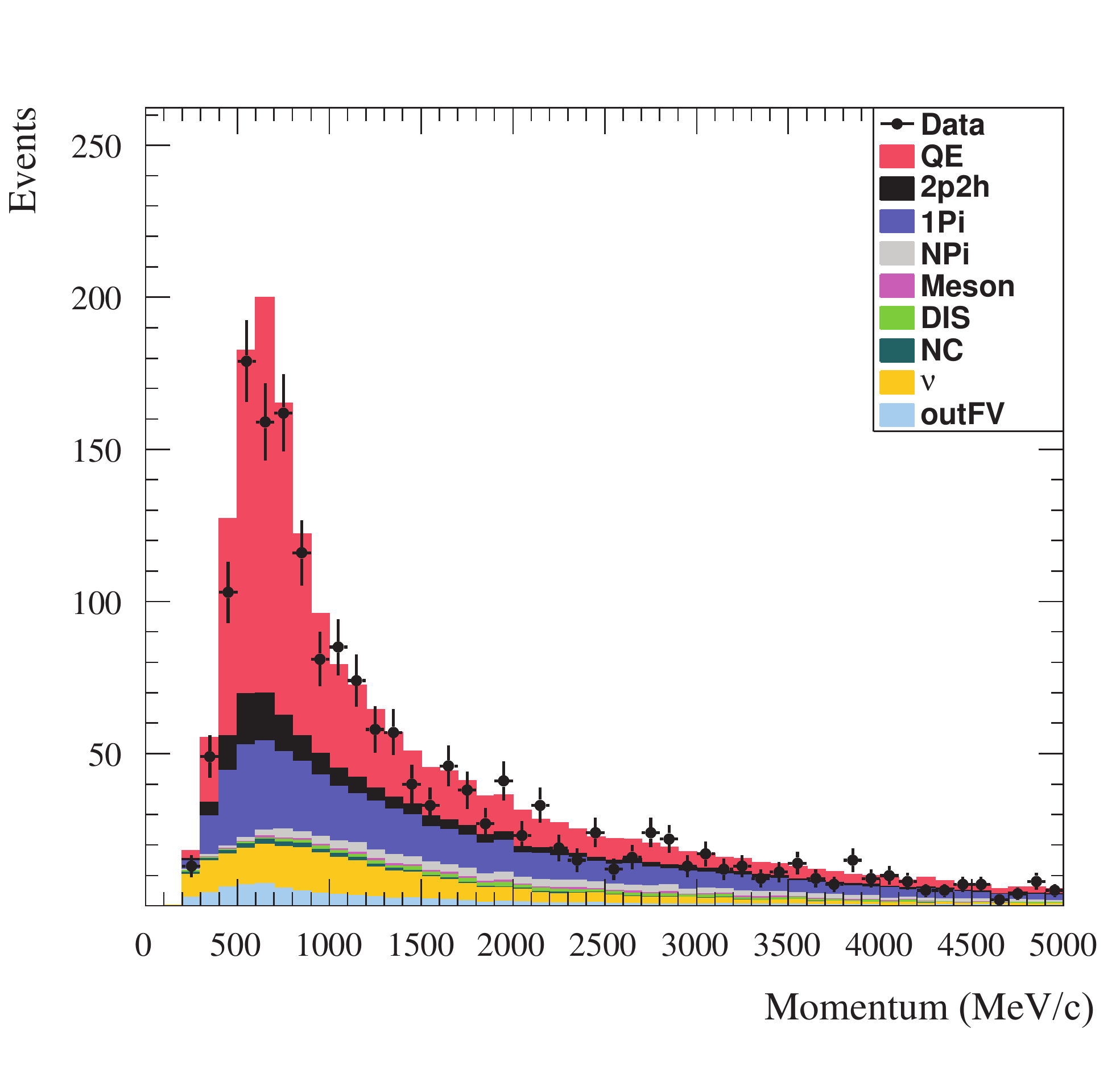}\includegraphics[scale=0.29]{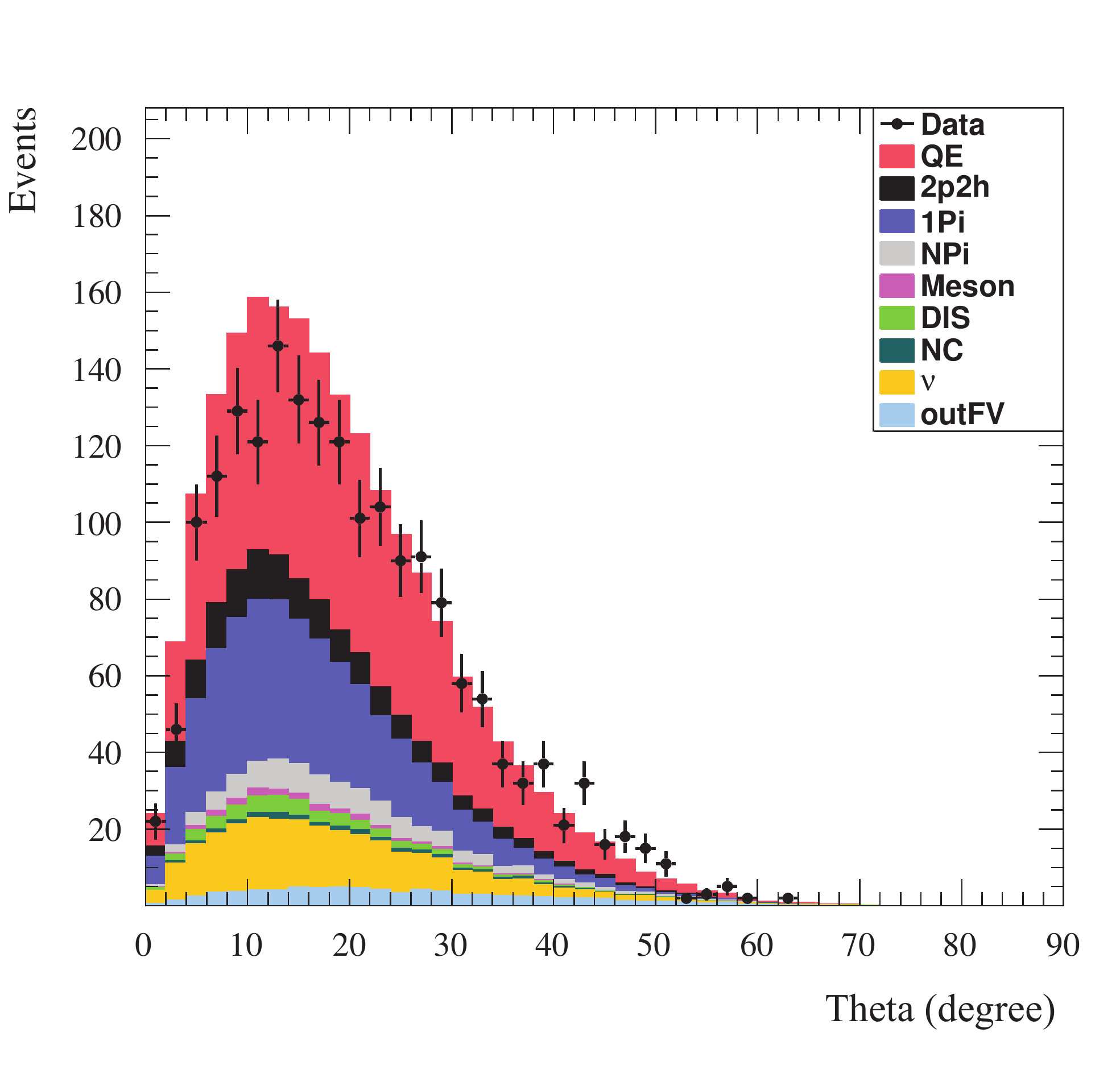}\includegraphics[scale=0.29]{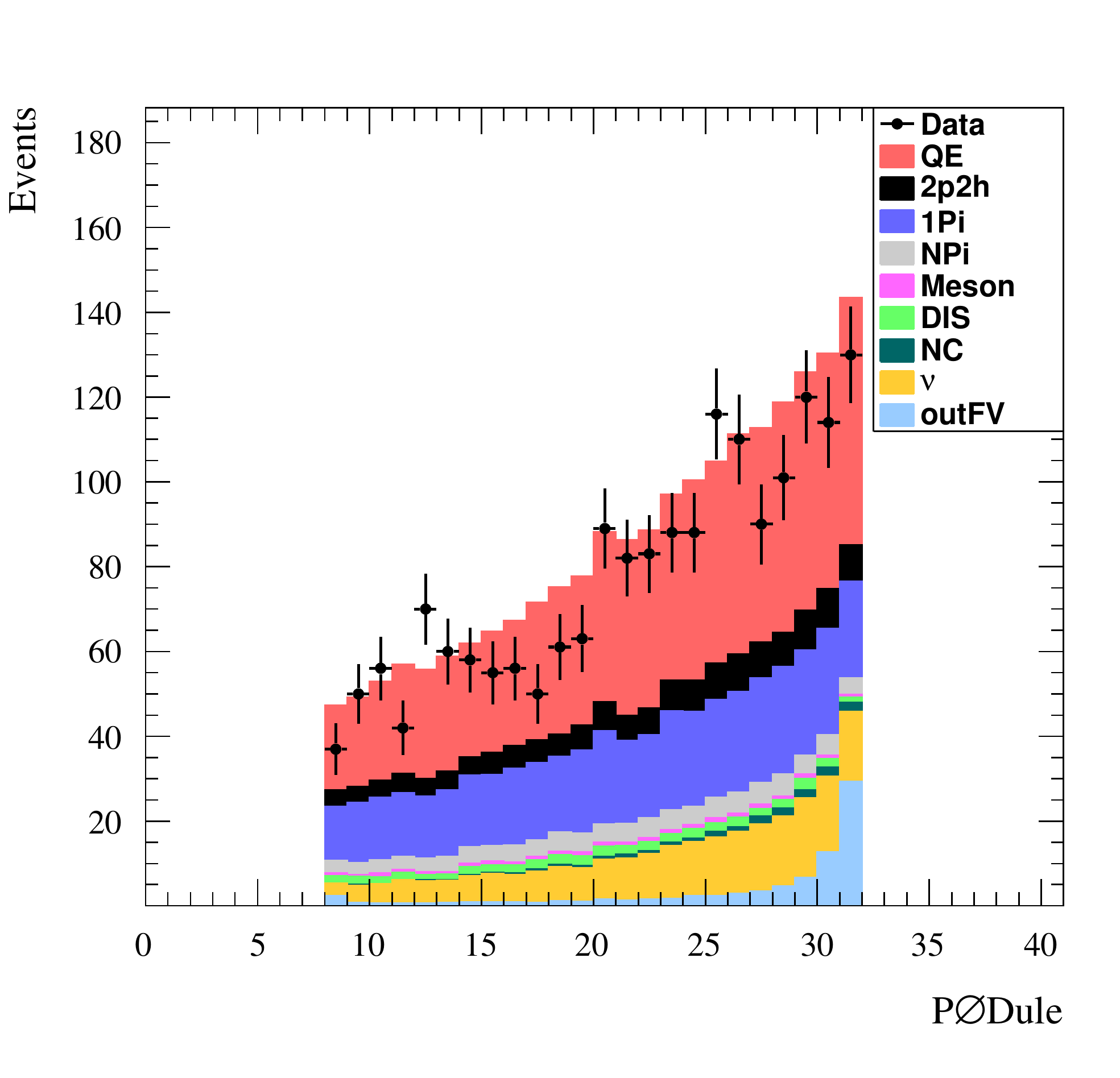}

\caption{\label{fig:RHC-antineutrino-beam}RHC beam CCINC \nubar
event candidate distributions of the $\mu^{+}$ momentum in MeV/c
(Left), the muon $\theta_{\mu}$ angle in degrees (Middle), and interaction
vertex position by P$\emptyset$Dule (Right). 
Note backgrounds in the
CCINC sample are the NC (dark green), $\nu_\mu$ induced events (yellow)
and the out of fiducial volume events (light blue). The $\nu_\mu$ backgrounds
in the RHC beam sample are much larger than the analogous $\bar\nu_\mu$ backgrounds
in the FHC beam sample.}
\end{figure*}

After independent reconstructions in the TPC and in the P$\emptyset$D, the analysis
uses an algorithm to match a 3D P$\emptyset$D track ending
near the most downstream edge of the P$\emptyset$D to a TPC track beginning
near the most upstream edge of the TPC.

The event selection is the following: 
\begin{enumerate}
\item The first requirement is good data quality for the data run.
After ND280 data is processed, the sub-detectors are evaluated 
run by run for
good timing with respect to the beam and 
checked to satisfy good detector calibrations.
Events are used only if their run passed data quality checks.
For each FHC (RHC) 
beam bunch there must be a negative (positive) TPC track
that is identified within $\pm$70~ns around the
nominal beam bunch time. 

\item A veto is applied to reject events whose vertex originated
outside the fiducial region but had a secondary interaction
inside the fiducial region. Also events with
single tracks that are broken into two tracks by the track reconstruction
are rejected. The event vertex is defined by the most upstream P$\emptyset$Dule
hit in the track. The vertex X-Y position is defined by the X-Y triangular
scintillator bars and the vertex Z position of the P$\emptyset$Dule. The fiducial
volume requires the vertex to be within -836 mm\ $<X<$ 864 mm and
-871 mm\ $<Y<$ 869 mm and inside one of the middle 24 P$\emptyset$Dules. 
The X boundaries are $\sim 250$ mm and the Y boundaries are
$\sim 236$ mm away from the ends of the X and Y scintillator bars, respectively.
\item The vertex must be in the
P$\emptyset$D water target fiducial volume. The charge is determined by the
curvature of the TPC track. Of all TPC tracks meeting these criteria, the
one with the highest reconstructed momentum at the start of the track
is chosen to be the lepton candidate. 
\item The RHC mode selection has an additional requirement
that the lepton track candidate is positively charged
and has the highest momentum of all 
charged tracks in the bunch.
\end{enumerate}

Due to the limited geometric acceptance of requiring a CC neutrino event vertex in the  P$\emptyset$D with
its muon track detected in the TPC, this analysis is inherently not sensitive to the
entire muon kinematic phase space. For this reason, we define a restricted
phase space, described in the next sub-section, that will cover the
part of the kinematic phase space where we have good acceptance.
Events that are reconstructed to have muon kinematics outside of the
restricted phase space will be rejected.
\begin{table}

\caption{
The fractional distributions of true MC interactions for selected events
defined at the initial interaction vertex according to the NEUT generator for the 
FHC beam (Left) and RHC beam (Right) modes. See text for descriptions
of each MC channel.\label{tab:The-fractional-distribution-1}}

\begin{tabular}{lc}
\hline 
\hline
\multicolumn{2}{c}{FHC beam}\\ 
\hline 
Mode & Fraction \\
\hline 
QE & 37.83\% \\ 
2p2h & 3.30\% \\ 
1Pi & 29.73\% \\ 
NPi & 11.01\% \\
Meson & 1.71\% \\
DIS & 11.27\% \\
NC & 1.50\% \\
$\overline{\nu}_{\mu}$ & 0.33\% \\
outFV & 3.32\% \\
\hline 
\hline
\end{tabular}\ %
\begin{tabular}{lc}
\hline 
\hline
\multicolumn{2}{c}{RHC beam}\tabularnewline
\hline 
Mode & Fraction\tabularnewline 
\hline 
QE & 47.27\%\tabularnewline
2p2h & 3.19\%\tabularnewline
1Pi & 24.14\%\tabularnewline
NPi & 5.05\%\tabularnewline
Meson & 1.04\%\tabularnewline
DIS & 2.32\%\tabularnewline 
NC & 0.99\%\tabularnewline 
$\nu_{\mu}$ & 11.93\%\tabularnewline 
outFV & 4.05\%\tabularnewline
\hline
\hline 
\end{tabular} 
\end{table}
For the FHC mode selection, 19,259 events are selected in data.
The number of selected events in the corresponding MC sample,
scaled to the same data PoT exposure is 19,566. In RHC mode,
1,869 events are selected in data and the scaled MC sample has 1,953 events. 
The muon $p$ and $\theta$ distributions for data events with MC predictions 
are shown for both modes
in  Figs. \ref{fig:FHC-Neutrino-beam} (Left and Middle) and
\ref{fig:RHC-antineutrino-beam} (Left and Middle), respectively. The plots 
include colored stacked histograms 
of MC interaction types to graphically display the composition
of the selected events. 

The fractional NEUT interaction types for the FHC and the
RHC beam modes are given in Table \ref{tab:The-fractional-distribution-1}
for the selected events described in Section IV. The MC channels defined\citep{joe-sam}
at the
initial interaction vertex according to NEUT are 
CCQE (QE), 2p2h, CC with 1 charged pion (1Pi), CC with
>1 charged pion (NPi), CC with K or $\eta$ meson (Meson), deep inelastic
scattering (DIS), neutral current (NC),
neutrino or antineutrino interaction
($\nu$ or $\overline{\nu}$), and events whose true vertex position was
outside the fiducial volume (outFV) region of the P$\emptyset$D . The 
resulting selected events, according to the MC simulation, 
are predominately CCQE, followed by CC events with 1 pion. 
Due to a substantial
$\nu_\mu$ flux contamination in the RHC beam and a large $\nu_\mu$ cross section,
the \nubar candidate sample has a larger background fraction (see yellow band
in Fig. \ref{fig:RHC-antineutrino-beam})
compared to the $\bar{\nu}_\mu$ background events in the FHC
beam sample. 
The $\nu_\mu$ in the
RHC beam flux is seen in Fig.\ref{fig:The-neutrino-FHC} (Bottom). 
The outFV backgrounds are roughly the same
fraction in both FHC and RHC beam samples. The selection 
produces a CCINC $\nu_\mu$ candidate event sample that is 94.8\% pure 
and a CCINC $\bar\nu_\mu$ cadidate event sample that is 83.0\% pure. 
The outFV backgrounds cluster in
the light blue bands in Figs. 
\ref{fig:FHC-Neutrino-beam} (Right) 
and
\ref{fig:RHC-antineutrino-beam} (Right) 
in the downstream P$\emptyset$Dules. These backgrounds are events whose 
vertices are outside and downstream of the fiducial volume but with 
an interaction that has a backwards going
track that enters the fiducial volume.

Additional checks between the data and MC event selections were performed by
comparing the event rates of vertices by detector P$\emptyset$Dule between data
and normalized selected MC events. The event rates by P$\emptyset$Dule are shown
for \nuandnubar
in the Figs. \ref{fig:FHC-Neutrino-beam} (Right) and \ref{fig:RHC-antineutrino-beam}
(Right), respectively. There is very good
agreement within statistics between the data and MC distributions, except the
momentum distribution in the FHC beam sample where the data is
1-2 sigma below the MC predictions near 0.6 GeV/c. 
The efficiency for the \nuandnubar events varies as a function of P$\emptyset$Dule.
Since the event selection requires a vertex in a P$\emptyset$Dule with a muon track
reconstructed in the TPC, the downstream P$\emptyset$Dules have a higher
efficiency than the upstream P$\emptyset$Dules. The events with vertices
in the more upstream P$\emptyset$Dule have smaller angular acceptance for
a muon track to pass through the TPC and the muon track will incur
more energy loss since it must pass through more P$\emptyset$Dules to reach
the TPC where it must be reconstructed. The $\nu$ event selection
efficiency in Fig. \ref{fig:FHC-Neutrino-beam} (right) from upstream
to downstream P$\emptyset$Dule varies from 37\% to 57\% whereas the \nubar
event selection efficiency Fig. \ref{fig:RHC-antineutrino-beam} (right)
varies from 39\% to 68\%. 

\begin{figure*}
\includegraphics[scale=0.35]{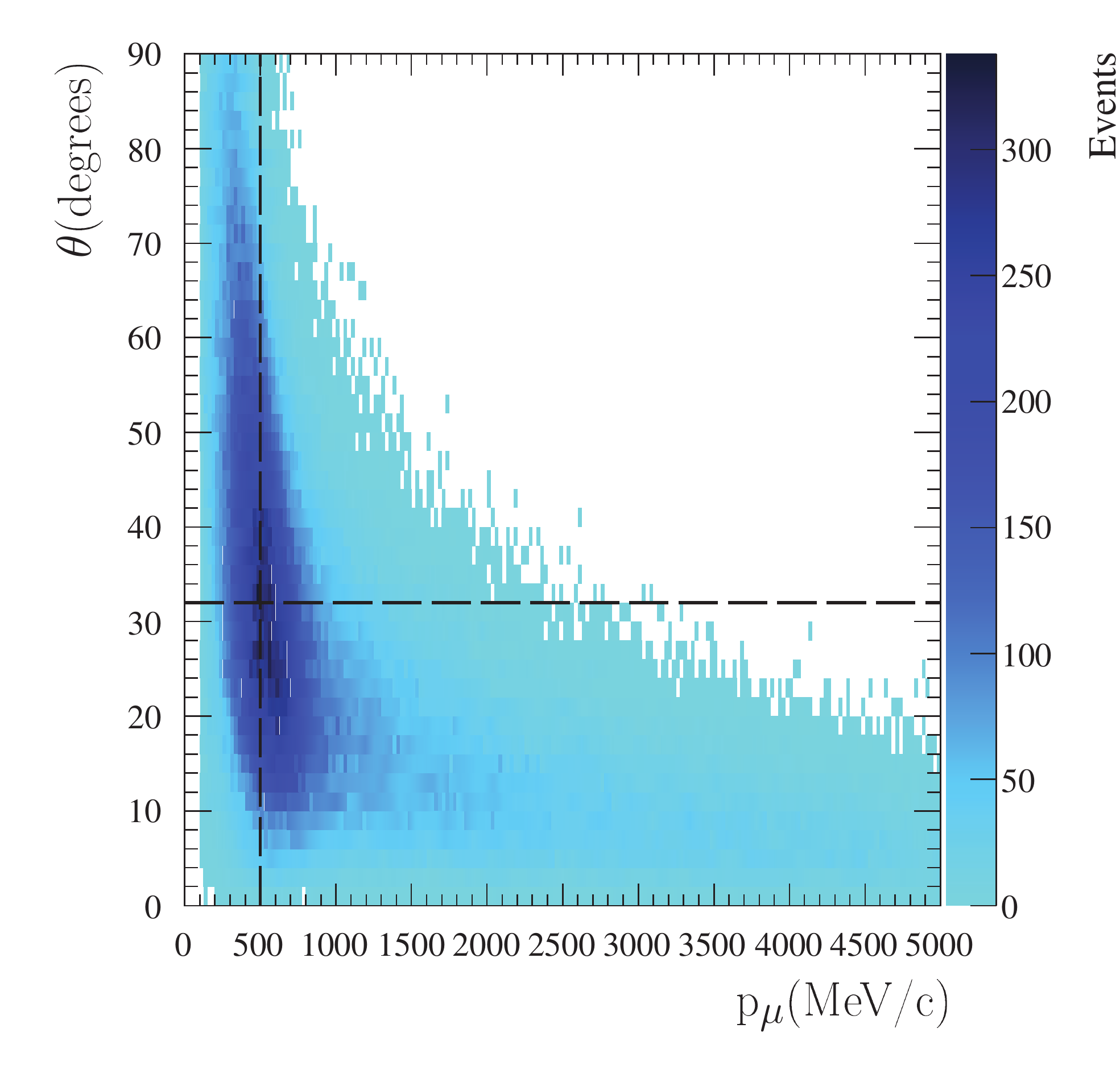}
\includegraphics[scale=0.35]{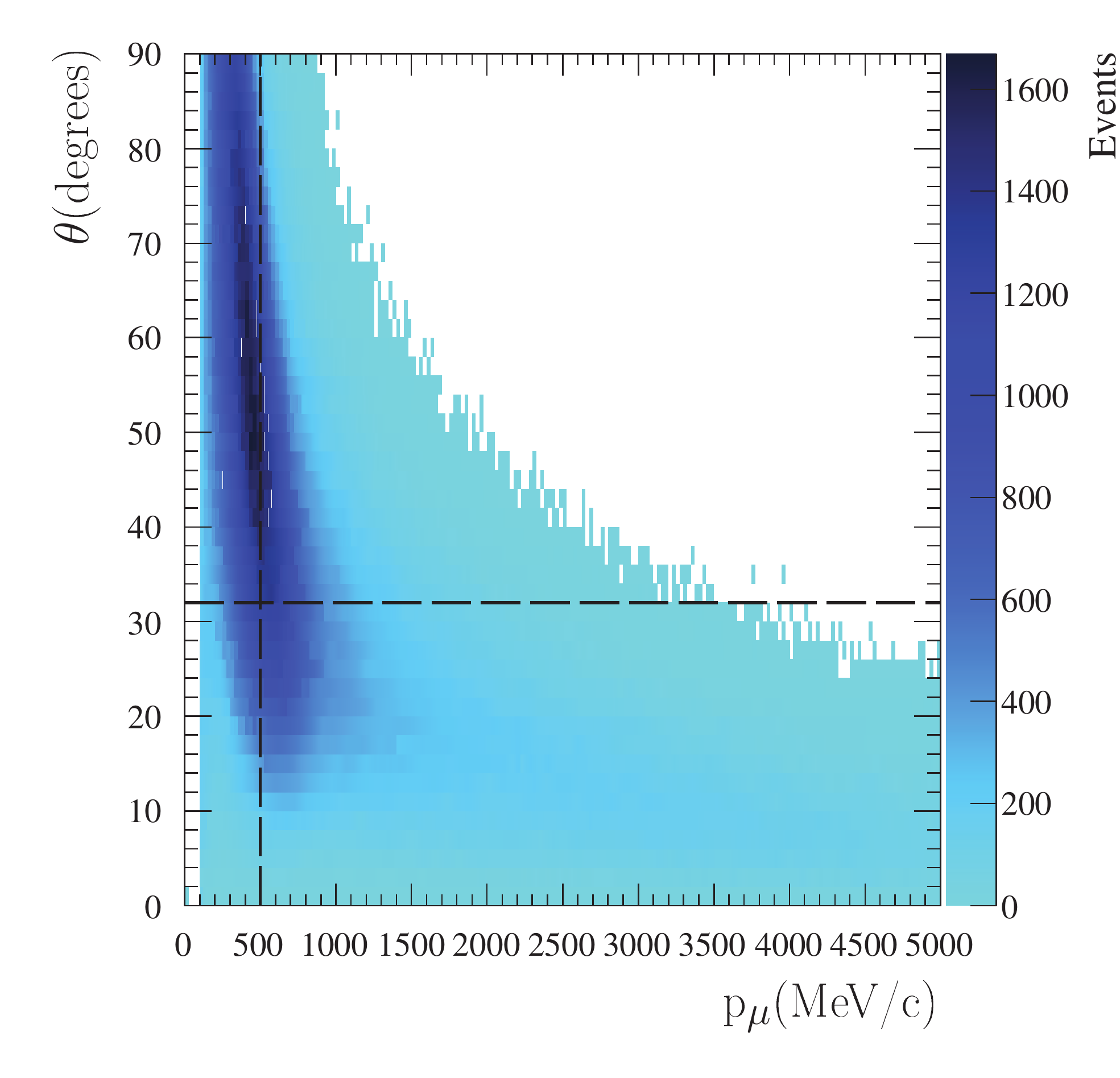}
\caption{\label{fig:full-CCinc-generated}
Left (Right) 2-D plots of  $\theta_{\mu}$ versus $p_{\mu}$
for RHC (FHC) beam events of $\mu^{+}\left(\mu^{-}\right)$
tracks using MC generated CCINC with full acceptance. The vertical
and horizontal solid lines correspond to 
$\theta_{\mu}=32\protect\textdegree$
and $p_{\mu}=500$ MeV/c, respectively. 
} 
\end{figure*}

\begin{figure*}
\includegraphics[scale=0.35]{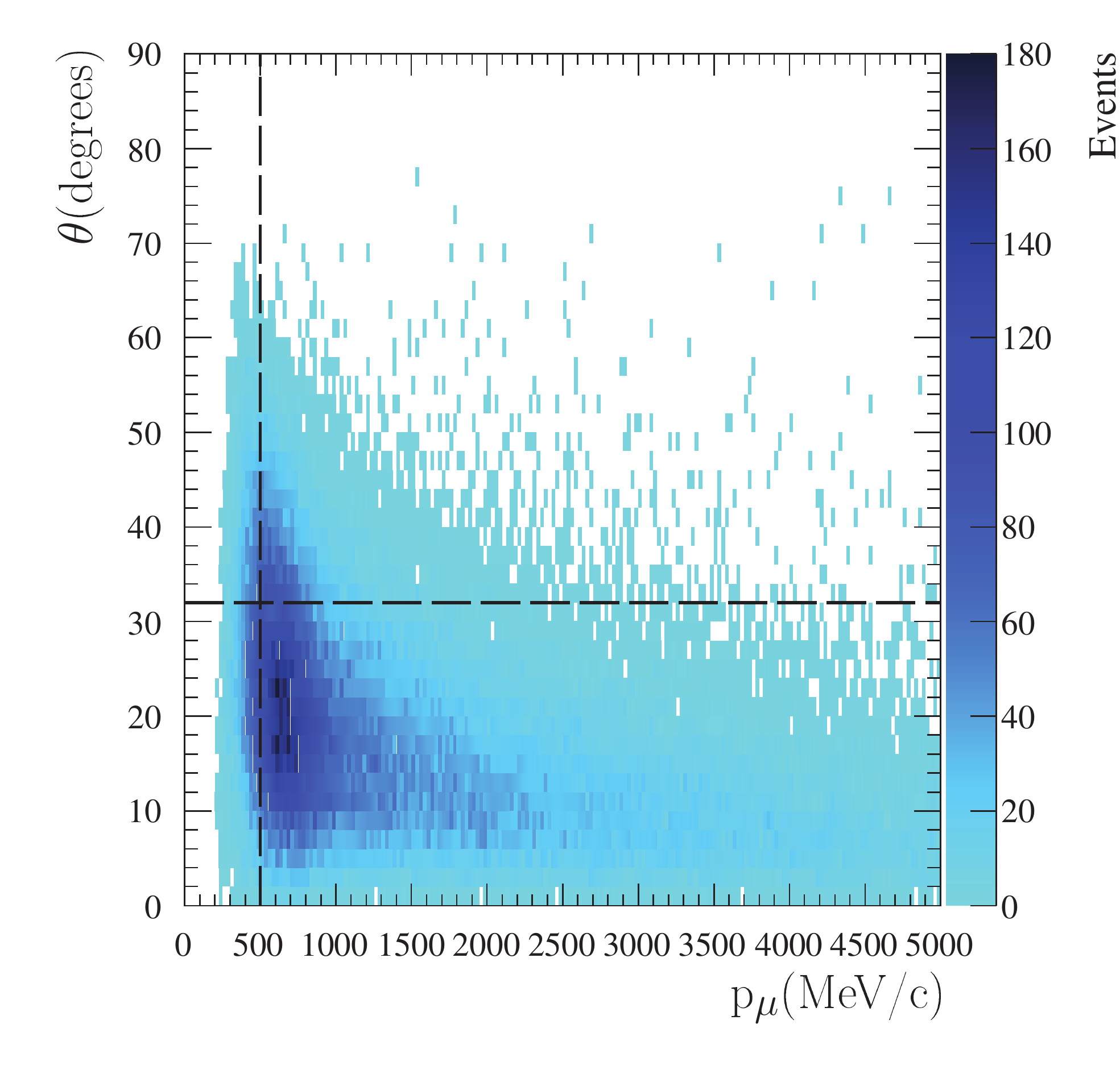}
\includegraphics[scale=0.35]{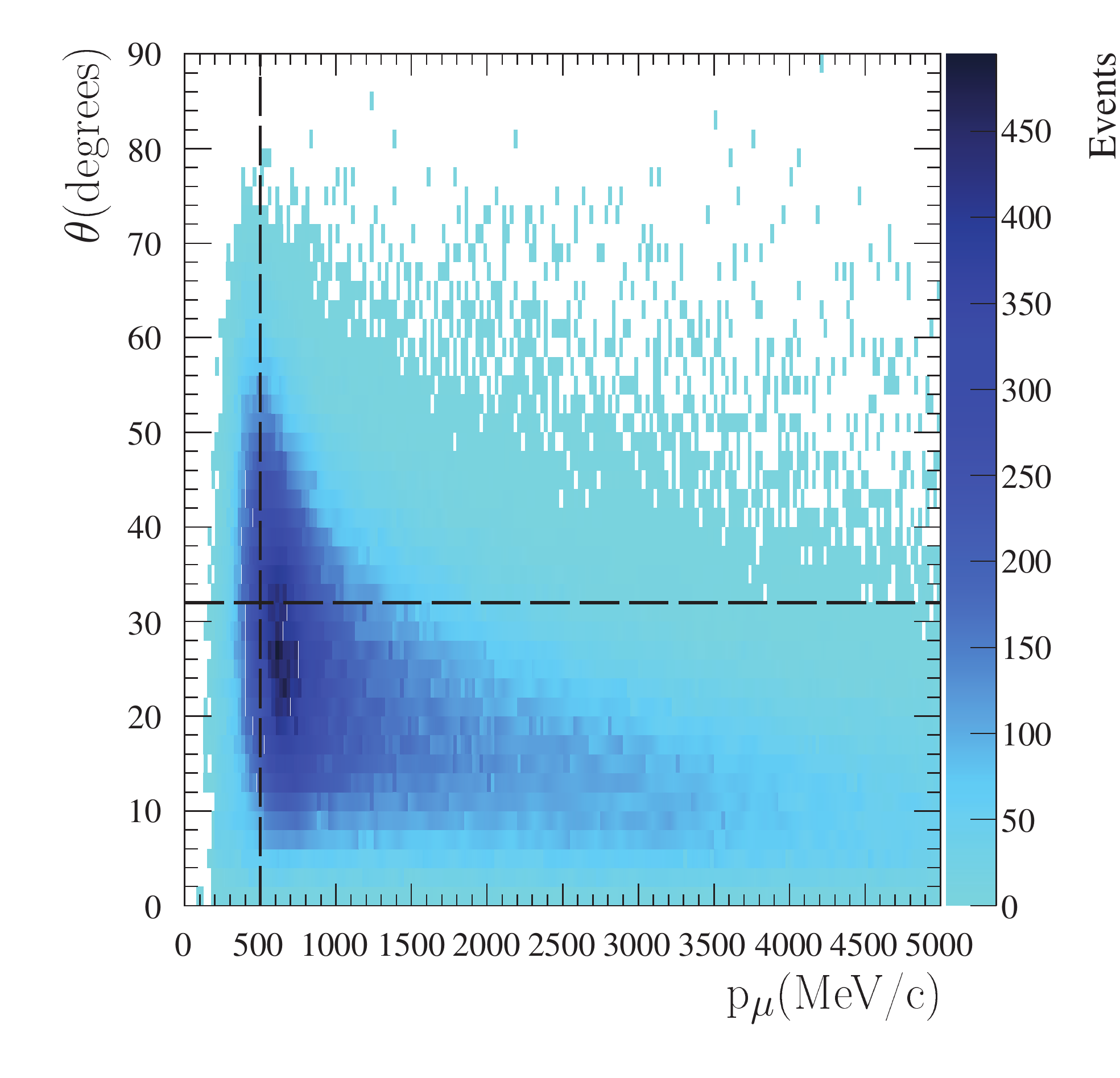}

\caption{\label{fig:P0D-vertex-and} Left (Right) 2-D plots of  $\theta_\mu$ versus $p_\mu$ 
for RHC (FHC) beam events of $\mu^+\left(\mu^-\right)$
tracks using MC generated CCINC that have a reconstructed P$\emptyset$D vertex and TPC muon
track. The vertical and horizontal solid lines correspond to 
$\theta_\mu=32\protect\textdegree$
and $p_\mu=500$ MeV/c, respectively. The restricted phase space
cut selection applies to events inside the
lower right
rectangular region 
defined by the dashed lines. }
\end{figure*}

\subsection{Kinematic selection}

The selected events for the RHC (FHC) samples require a vertex in
the P$\emptyset$D and a $\mu^{+}\left(\mu^{-}\right)$ reconstructed track in
the TPC detector. 
This limits or restricts the available kinematic
phase space of the CCINC events such that
certain kinematic regions are not measured. 
These unmeasured regions in the laboratory frame
have low muon momentum 
$p_\mu<500$ MeV/c or large muon polar angles
$\theta_{\mu}>32\textdegree$.

These kinematic boundaries are displayed in Figs. \ref{fig:full-CCinc-generated}
and \ref{fig:P0D-vertex-and} Left (Right)
where the  $\theta_{\mu}$ versus $p_{\mu}$
2-D plots are shown for the RHC (FHC) samples. In Fig.
\ref{fig:full-CCinc-generated} Left (Right) are the generated MC
full acceptance CCINC events for the RHC (FHC) samples. The $\nu_\mu$
mode has more events with larger $\theta_{\mu}$ polar angles since
the $\mu^{-}$ angular distribution is more isotropic than the $\mu^{+}$
in the \nubar mode whose muon tracks are more forward. 
In Fig. \ref{fig:P0D-vertex-and} Left (Right)
are the generated MC CCINC events that have a P$\emptyset$D vertex and a $\mu^+$ ($\mu^-$)
track reconstructed in the TPC for the RHC (FHC) samples. The regions
below horizontal lines  where $\theta_{\mu}<32\textdegree$ and right of
the vertical dash lines where $p_\mu>500$ MeV/c are detector regions
that have non-zero acceptance
and reconstructed events for both the
FHC and the RHC samples.
Hence we use these two kinematic
restrictions in the cross section measurements.
%
%
The resulting 
reconstructed
restricted phase space selection in
the $\nu_\mu$ mode has 14,398 data events and 
a corresponding MC sample, scaled to the same
data PoT exposure, contains 15,284 events. In the \nubar
mode, 1,461 data events are selected and a scaled MC sample
has 1,634 events. From a study of MC truth selected events, this restricted
phase space selection changed the mean value of 
neutrino energies below 2 GeV in the FHC sample from
0.83 GeV (unrestricted) to 1.14 GeV (restricted) 
and in the RHC sample from 0.84 GeV (unrestricted) to 1.08 GeV (restricted). 
In addition, the $\nu_\mu$ and  \nubar MC samples
contained 2.19\% and 1.33\% events, respectively, 
whose true kinematic value was outside the restricted phase
space region, but its reconstructed value migrated to be 
inside the restricted phase space region.
These events are kinematic backgrounds that originated
from the same physics process.

%% file: Sections/AnalysisMethods-2.tex
\section{Analysis Methods}

The number of neutrino interactions in the fiducial volume of the P$\emptyset$D, $N_{signal}$,
can be expressed as the product of the signal cross section per target, $\sigma$, the number of targets, $N_{targets}$, and the integrated flux, $\Phi$, of incident
neutrinos per unit area, as
\begin{equation}
N_{signal}=\sigma N_{targets}\Phi.\label{eq:Analysis:1-1}
\end{equation}
Hence the cross section becomes
\begin{equation}
\sigma=\frac{N_{signal}}{\Phi N_{targets}}.\label{eq:Analysis:2-1}
\end{equation}
Using our event selection on data, we obtain a candidate signal event sample in our fiducial volume.
This process is not 100\% efficient and also some non-signal (background) events 
are included.
To account for this, the MC simulation is used to estimate in our sample
the number of background events and the number of
signal events.  The backgrounds from the FHC (RHC) beam samples
include non-CCINC events from the neutrino (antineutrino) beam as well as
events created from the antineutrino (neutrino) flux.  
In addition, the MC simulation generates
total number of signal events that were produced. 
If the rate of 
restricted phase space
selected data events is $N_{selected}^{data}$
and the predicted number of selected background events is $B^{MC}$,
the observed number of signal 
candidates 
in our fiducial volume is
\begin{equation}
N_{selected\ signal}=N_{selected}^{data}-B^{MC}\label{eq:Analysis:3-1}
\end{equation}
which include migration events.
Next we redefine the selection efficiency
$\epsilon$ as
\begin{equation}
\epsilon=\frac{N_{selected\ signal}^{MC}}{N_{generated\ signal}^{MC}}.\label{eq:Analysis:4-1}
\end{equation}
where the $N^{MC}_{selected\ signal}$ is the number of signal candidates whose reconstructed kinematics 
are in the restricted phase space and  $N^{MC}_{generated\ signal}$ is the total number of generated 
signal events whose true kinematics are in the restricted phase space. We note that
 $N^{MC}_{selected\ signal}$ includes a small fraction of migration events
as described at the end of Section IV.B.
With these definitions, the 
restricted phase space
signal event rate is
\begin{equation}
N_{signal} = \frac{N_{selected}^{data}-B^{MC}}{\epsilon}.\label{eq:Analysis:5-1}
\end{equation}
In Eqn.(\ref{eq:Analysis:5-1}) the numerator
is the number of signal candidates whose reconstructed kinematics are in the restricted 
phase space, and this is combined with the
denominator  $\epsilon$ from Eqn.(\ref{eq:Analysis:4-1})
to give the proper estimate 
of $N_{signal}$ that represents the number of signal events 
whose kinematics are in the true restricted phase space.
The neutrino cross section is
\begin{equation}
\sigma (\nu_{\mu})=\frac{N_{selected}^{data}-B^{MC}}{\epsilon N_{targets}\Phi}.\label{eq:Analysis:6-1}
\end{equation}

In addition to the cross sections given above, the measured ratio of cross sections $R(\nu,\overline{\nu})$ and rates $r(\nu,\overline{\nu})$
are defined as
\begin{equation}
R (\nu,\overline{\nu}) \equiv \frac{\sigma(\overline{\nu}_{\mu})}{\sigma(\nu_{\mu})}=\frac{\overline{N}_{selected}^{data}-\overline{B}^{MC}}{N_{selected}^{data}-B^{MC}}\times\frac{\overline{\epsilon}}{\epsilon}\times\frac{\varPhi}{\overline{\varPhi}}\label{eq:Analysis:7-1}
\end{equation}
and
\begin{equation}
r (\nu,\overline{\nu}) \equiv \frac{n(\overline{\nu}_{\mu})}{n(\nu_{\mu})}=\frac{\overline{N}_{selected}^{data}-\overline{B}^{MC}}{N_{selected}^{data}-B^{MC}}\times\frac{\overline{\epsilon}}{\epsilon}.\label{eq:Analysis:7-2}
\end{equation}
The overlined quantities are obtained from the antineutrino
selections as described above and those without overlines represent the
neutrino mode selection.  Finally, other observables are introduced and defined; 
the sum $\Sigma(\nu,\overline{\nu})$, 
difference $\Delta(\nu,\overline{\nu})$, 
and asymmetry $A(\nu,\overline{\nu})$
formed from the $\nu_\mu$ and \nubar 
cross sections, as
\begin{equation}
\Sigma(\nu,\overline{\nu}) \equiv \sigma(\nu_{\mu})+\sigma(\overline{\nu}_{\mu}),\label{eq:Analysis:sum}
\end{equation}
\begin{equation}
\Delta(\nu,\overline{\nu}) \equiv \sigma(\nu_{\mu})-\sigma(\overline{\nu}_{\mu})\label{eq:Analysis:diff}
\end{equation}
and
\begin{equation}
A(\nu,\overline{\nu}) \equiv 
\frac{\sigma(\nu_{\mu})-\sigma(\overline{\nu}_{\mu})}{\sigma(\nu_{\mu})+\sigma(\overline{\nu}_{\mu})}.\label{eq:Analysis:asym}
\end{equation}

%% file: Sections/PropagationOfSystematics-b.tex
\section{Cross Section and Ratio Systematic Errors}
The systematic errors on cross sections and ratios of cross sections 
in this analysis are due to 
uncertainties on the number of selected background events, the incident 
neutrino flux, the number of targets in the detector, and the selection efficiencies. 
The sources of systematic uncertainties can be categorized into three groups:
beam flux prediction, 
neutrino and antineutrino interaction models 
and detector response.
The largest source of uncertainty is due to the beam flux.
\subsection{Beam flux uncertainty}
The beam flux uncertainty sources can be separated
into two categories: uncertainties 
of the hadronic interactions, in the graphite target and reinteractions in the
horn, and
T2K beamline inaccuracies. \\

The beam flux uncertainty is dominated by the
uncertainty on the modeling of the hadron interactions, including
uncertainties on the total proton-nucleus production cross
section, pion and kaon multiplicities, and secondary nucleon production. \\

The hadronic interactions in the target where the primary
proton beam first interacts and produces the majority of
the secondary pions is simulated by the FLUKA2011 package
which creates MC neutrino and antineutrino flux samples.
Uncertainties on the proton beam properties, horn current,
hadron production model and alignment are taken into
account to produce an energy-dependent systematic uncertainty
on the neutrino flux.
These uncertainties are propagated to the T2K neutrino beam flux
prediction by reweighting MC flux samples. 
The total proton-nucleus production cross section uncertainty
is adjusted to replicate discrepancies between NA61/SHINE measurements
and other external data sets. 
\\

The flux smearing is done using toy MC data sets that are
based on the FHC and RHC beam flux uncertainty covariance matrices. 
The resulting $\pm1\sigma$ change in the cross section is taken as
the systematic error associated with the beam flux. 
These uncertainties on individual cross sections
lead to 9\% errors whereas the errors on the ratio are 4\%
due to correlated neutrino and antineutrino flux covariance errors. 
Table III summarizes the systematic errors due to the beam flux 
uncertainties on the cross sections and combinations of cross sections.
These results have been cross checked with analytic calculations.
The fractional errors on ratios have smaller errors due to
cancellations of correlated errors between the
neutrino and antineutrino modes.
\begin{table}
\caption{
\label{tab:BeamFluxSummary}
Summary table for one standard deviation errors due to beam flux uncertainties. (fractionl errors in $\%$)
}
\resizebox{\columnwidth}{!}{
\begin{ruledtabular}
\begin{tabular}{cccccc}
$\sigma(\overline{\nu})$ & $\sigma(\nu)$ & $R (\nu,\overline{\nu})$ 
& $A(\nu,\overline{\nu})$ & $\Sigma(\nu,\overline{\nu})$ & $\Delta (\nu,\overline{\nu})$ \\
\hline
 $\pm9.37$ & $\pm9.14$ & $\pm3.58$ & $\pm3.35$ & $\pm9.17$ & $\pm9.42$ \\
\end{tabular}
\end{ruledtabular}
}
\end{table}


\subsection{Interaction model uncertainty}

The interaction model uncertainties were calculated by a data-driven method \citep{int-model} 
where the NEUT predictions were compared 
to external neutrino-nucleus data in the energy region relevant for T2K. 
Some of the NEUT model parameters are fitted and 
assigned mean and $1$ $\sigma$ error values that allow
for differences between NEUT 
and the external data. \\


\begin{table*}
\caption{
\label{tab:PhysicsModelPiFSISummary}
Summary table for physics model uncertainties for restricted
phase space measurements (fractional errors in $\%$).
}

\begin{tabular}{lcccccc}
\hline
\hline
Parameter & 
$\sigma(\overline{\nu})$ & $\sigma(\nu)$ & $R (\nu,\overline{\nu})$ & $A(\nu,\overline{\nu})$ & $\Sigma(\nu,\overline{\nu})$ & $\Delta (\nu,\overline{\nu})$ \\
\hline
$M^{QE}_A$ 
& $\pm0.51$ & $\pm0.14$ & $\pm0.37$ & $\pm0.32$ & $\pm0.24$ & $\pm0.08$ \\
$p_F\ (^{12}$C) 
& $\pm0.01$ & $\pm0.02$ & $\pm0.01$ & $\pm0.01$ & $\pm0.02$ & $\pm0.02$ \\
$p_F\ (^{16}$O) 
& $0$ & $\pm0.01$ & $0$ & $0$ & $\pm0.01$ & $\pm0.01$ \\
MEC norm\ $(^{12}$C) 
& $\pm0.30$ & $\pm0.44$ & $\pm0.14$ & $\pm0.12$ & $\pm0.40$ & $\pm0.52$ \\
MEC norm\ $(^{16}$O) 
& $\pm0.18$ & $\pm0.24$ & $\pm0.06$ & $\pm0.05$ & $\pm0.22$ & $\pm0.27$ \\
$E_B\ (^{12}$C) 
& $\pm0.01$ & $\pm0.01$ & $\pm0.02$ & $\pm0.02$ & $0$ & $\pm0.02$ \\
$E_B\ (^{16}$O) 
& $\pm0.01$ & $\pm0.01$ & $\pm0.02$ & $\pm0.02$ & $0$ & $\pm0.02$ \\
$C_5^A$(0) 
& $\pm0.70$ & $\pm0.46$ & $\pm0.24$ & $\pm0.21$ & $\pm0.53$ & $\pm0.32$ \\
$M_A^{1\pi}$ 
& $\pm0.99$ & $\pm0.28$ & $\pm0.75$ & $\pm0.65$ & $\pm0.44$ & $\pm0.21$ \\
$I=\frac{1}{2}$ Bkg 
& $\pm0.29$ & $\pm0.21$ & $\pm0.08$ & $\pm0.07$ & $\pm0.23$ & $\pm0.17$ \\
$\nu_e/\nu_{\mu}$
& $\pm0.02$ & $0$ & $\pm0.01$ & $\pm0.01$ & $\pm0.01$ & $0$   \\
CC Other shape 
& $\pm0.65$ & $\pm0.70$ & $\pm0.06$ & $\pm0.79$ & $\pm0.06$ & $\pm0.75$ \\
CC Coherent 
& $\pm0.01$ & $\pm0.01$ & $0$ & $\pm0.05$ & $\pm0.69$ & $\pm0.73$ \\
NC Coherent 
& $0$ & $0$ & $0$ & $0$ & $0$ & $0$ \\
NC Other 
& $\pm1.28$ & $\pm0.39$ & $\pm0.89$ & $\pm0.77$ & $\pm0.63$ & $\pm0.14$ \\
$\pi$ FSI
& $\pm0.16$ & $\pm0.19$ & $\pm0.11$ & $\pm0.09$ & $\pm0.18$ & $\pm0.23$ \\
MEC norm Other & $\pm0.08$ & $\pm0.15$ & $\pm0.07$ & $\pm0.20$ & $\pm0.13$ & $\pm0.20$ \\
\hline
Total & $\pm2.13$ & $\pm1.16$ & $\pm1.56$ & $\pm1.36$ & $\pm1.31$ & $\pm1.32$\\
\hline
\hline
\end{tabular}
\end{table*}

\begin{table*}
\caption{
\label{tab:DetectorResponseSummary} 
Summary table for detector response uncertainties (fractional errors in $\%$).
}

\begin{tabular}{lcccccc}
\hline
\hline
Parameter &
$\sigma(\overline{\nu})$ & $\sigma(\nu)$ & $R (\nu,\overline{\nu})$ & $A(\nu,\overline{\nu})$ & $\Sigma(\nu,\overline{\nu})$ & $\Delta (\nu,\overline{\nu})$ \\
\hline
TPC tracking Efficiency
& $\pm0.37$ & $\pm0.32$ & $\pm0.04$ & $\pm0.04$ & $\pm0.34$ & $\pm0.29$ \\
Charge misidentification
& $\pm0.37$ & $\pm0.32$ & $\pm0.04$ & $\pm0.04$ & $\pm0.34$ & $\pm0.29$ \\
Sand/Rock muon interference
& $\pm1.45$ & $\pm2.20$ & $\pm0.74$ & $\pm0.70$ & $\pm1.99$ & $\pm2.70$ \\
Fiducial mass
& $\pm0.96$ & $\pm0.96$ & $0$ & $0$ & $\pm1.36$ & $\pm1.36$ \\
Fiducial volume boundaries
& $\pm0.13$ & $\pm0.97$ & $\pm0.83$ & $\pm1.39$ & $\pm0.77$ & $\pm0.74$ \\
\hline
Total & $\pm1.82$ & $\pm2.63$ & $\pm1.11$ & $\pm1.02$ & $\pm2.58$ & $\pm3.35$\\
\hline
\hline
\end{tabular}
\end{table*}

The CCQE model in NEUT is based on the Llewellyn-Smith neutrino-nucleon scattering model \citep{LSmith}
with a dipole axial form factor and the BBBA05 vector form factors \citep{RBradford}.
The NEUT generator uses the Smith-Moniz RFG model \citep{RSmith}
and includes an implementation of both the random phase approximation (RPA) correction \citep{JNieves}
and the 2p2h Nieves model \citep{JNieves}.
The NEUT resonant pion production is based on the Rein-Sehgal model \citep{DRein}
with updated form factors from Ref. \citep{KGraczyk}.
The DIS model
used in NEUT includes both the structure function from Ref. \citep{MGluck}
and the Bodek-Yang correction \citep{ABodek}.
The NEUT MC generator includes various model parameters to describe the different models, uncertainties and approximations. 
The axial mass $M_A^{QE}$ 
was set to 1.21 GeV/c$^2$ based on the Super-Kamiokande atmospheric data and the K2K data.
The 1$\sigma$ error on $M^{QE}_A$ was set to 0.41 GeV/c$^2$.
The large uncertainty on this parameter is due to the disagreements between
recent experimental measurements and bubble chamber results\citep{maqe}.
The Fermi gas momentum parameter ($p_F$) values and their errors are set to 
223 MeV/c and 225 MeV/c for Carbon and Oxygen respectively 
with both errors set to $\pm$12.7 MeV/c. 
The Fermi gas binding energy ($E_B$) parameter was set to 
25 MeV and 27 MeV for Carbon and Oxygen respectively 
with both errors set to $\pm$9 MeV. 
The Nieves model 2p2h normalization to  $1\pm1$ 
for both Carbon and Oxygen,
the resonant pion production model in NEUT used the Graczyk and Sobczyk 
form factors $C^A_5(0)$ 
and the $I=\frac{1}{2}$ background scale 
were set to $1.01\pm0.12$ and $1.20\pm0.20$ respectively. 
The nominal axial mass $M^{RES}_A$ was set to $0.95\pm0.15 {\rm\  GeV/c}^2$.
Additional uncertainties are $\nu_e/\nu_\mu$ cross section factor 
that was set to $1.00\pm0.02$. Both CC and NC coherent uncertainties 
based on the Rein-Sehgal model
were set to $1\pm1$ and $1.0\pm0.3$ respectively. 
Moreover, for CC and NC interactions, additional scale factors
were set to $0.0\pm0.4$ and $1.0\pm0.3$ respectively.
In addition the CC other is an energy dependent factor\citep{tpc-analysis-paper}
and the NC other is a normalization factor.
The $\pi$ Final State Interaction (FSI) uncertainties 
are tuned to a pion-nucleus scattering data, and 
other smaller corrections were included  \citep{int-model}.

Variation of model parameters within their errors ($\pm1\sigma$) 
was used to estimate their effect on the final observables in order to
determine final measurement uncertainties.
A summary of the parameters and their effects on the overall normalization 
are shown in Table \ref{tab:PhysicsModelPiFSISummary}.

\subsection{Detector response uncertainty}
The detector response uncertainty studies 
used data samples supported with MC samples
and measurements of the target weight.
The three dominant detector response 
systematic uncertainties 
are caused by the fiducial volume boundaries, 
the sand/rock muon interactions 
and the mass of the target in the fiducial volume.
There were small uncertainties from reconstruction and
charge misidentification from the TPC measurements. 
All the sources of detector response errors considered 
in the analysis are given 
in Table \ref{tab:DetectorResponseSummary}.

The fiducial volume systematics were
estimated by varying its  boundaries.
The sand/rock muon interactions occurring upstream and 
in the surrounding ND280 volume
could create tracks passing through the P$\emptyset$D and TPC detectors,
mimicking a CCINC event. 
Another source of detector systematics was the mass of the target 
in the fiducial volume. The uncertainty due to the fiducial mass 
was conservatively estimated to be 0.96\%
from the measured mass of the detector material during construction and
the water mass measured during filling the water bags.

%% file: Sections/Results-a-Erez_Jan24.tex
\section{Results}

\subsection{Cross sections and ratios}

The flux averaged cross section and ratio values measured in the FHC
and RHC samples are extracted from the flux, the number of targets,
MC efficiencies and MC background estimates. The input parameters
are given in Table \ref{tab:input to final cross sections} and the
results for the restricted (full) phase space selections are given in
Tables \ref{tab:Restricted-phase-space-cross_sections}( \ref{tab:Full-phase-space_cross_sections}).
The systematic errors in Table \ref{tab:Restricted-phase-space-cross_sections}
are determined by adding in quadrature the errors in 
Tables III, IV and V. For example the fractional R error,
taken from the three Tables, is 4\% and this yields 0.015
for the absolute systematic R error in Table VII. 

In Table \ref{tab:input to final cross sections},
for the restricted phase space results,
the input parameters
include the $\nu_\mu$($\overline\nu_\mu$) fluxes normalized to PoT in the FHC(RHC) samples. 
The number of nucleon targets is given for both the data and MC
which slightly differed. The number of reconstructed MC events is
given scaled to the equivalent data PoT. The data/MC generated corrected
events are defined as the reconstructed data/MC generated events,
minus the MC background and divided by the MC CCINC efficiencies.

In Table \ref{tab:Full-phase-space_cross_sections}, the full phase
space results are extrapolated
by scaling the restricted values in Table \ref{tab:input to final cross sections}
by the ratio of the total to restricted cross sections as predicted by the
NEUT MC generator.  The single errors combine the
statistical and systematic errors, which included 
model uncertainties on the assumed values of
$M_A^{QE}$ and the 2p2h $C^{12}$  and $O^{16}$ parameters in the scaling factor.
The errors
on the $\nu_\mu$ and \nubar cross sections due to these parameter
uncertainties
were assumed to be totally uncorrelated leading to a conservative
estimate of the systematic errors on the full phase space
ratio of cross sections.

The cross section calculations use Eqn.(\ref{eq:Analysis:6-1})
and the ratio $R(\nu,\overline{\nu})$ is obtained from Eqn.(\ref{eq:Analysis:7-1}) where
we note the number of targets drops out. 
We find $\approx 10\%$
systematic cross section
errors whereas the ratio of cross sections $R(\nu,\overline{\nu})$ error
has a factor $\times2$
smaller values of $4.0\%$ 
errors for the restricted 
phase space.
These systematic errors are mainly due to the flux uncertainties 
on the flux prediction which
have strong correlations between neutrino and antineutrino fluxes
which largely cancel in the ratio.  
The flux predictions for neutrino mode and antineutrino mode are correlated through measurements that are used as inputs to the flux calculation.  These measurements include the proton beam current measurement, the measurement of the primary proton interaction rate by NA61/SHINE, and the measurement of secondary particle interaction rates by other hadron interaction experiments.
The measured ratio of rates  $r(\nu,\overline{\nu})$  
given in Eqn.(\ref{eq:Analysis:7-2}) represents the ratio
of $\nu_\mu$ and $\bar\nu_\mu$ event rates which depends on the integrated
FHC and RHC flux and so its value depends on the particular experiment and data taking periods.
The event rate ratio $r(\nu,\overline{\nu})$  
fractional systematic uncertainty is the same as 
cross section ratio
$R(\nu,\overline{\nu})$, except
it does not include the flux errors given in Table \ref{tab:BeamFluxSummary}. The fractional 
systematic errors are $1.92\%$ 
for the restricted 
phase space selections.


\begin{table*}
\caption{\label{tab:input to final cross sections}
Tabulation of flux, targets,
and data/MC events used in the cross section calculations.
The data corrected values are background
subtracted and divided by the MC efficiency.
}
\begin{tabular}{lccc}
\hline
\hline
Inputs for Cross Sections & Units & RHC $\overline{\nu}$ mode & FHC $\nu$ mode \tabularnewline
\midrule 
Integrated flux  & {[}cm$^{2}$/10$^{21}$ PoT{]}  & 1.477$\times10^{13}$  & 1.823$\times10^{13}$\tabularnewline
Number of targets (data)  & {[}Nucleons{]}  & 3.147$\times10^{30}$  & 3.147$\times10^{30}$ \tabularnewline
Number of targets (MC)  & {[}Nucleons{]}  & 3.119$\times10^{30}$  & 3.119$\times10^{30}$ \tabularnewline
\midrule 
Number of data/MC events (restricted PS) & {[}Events{]} & 1,498/1,634 & 14,398/15,284
 \tabularnewline
Data corrected (restricted PS) & {[}Events/10$^{21}$ PoT{]}   & 41,821$\pm$1,334 & 138,576$\pm$1,249
\tabularnewline
\hline
\hline
\end{tabular}
\end{table*}

\begin{table*}
\caption{\label{tab:Restricted-phase-space-cross_sections} Restricted phase
space cross section and ratio final results.}

\begin{tabular}{crcl}
\hline
\hline 
\multicolumn{4}{c}{Cross Sections {[}$\times10^{-39}$cm$^{2}$/nucleon
{]} }\tabularnewline
\hline 
$\sigma\left(\overline{\nu}\right)$  & 0.900  & $\pm$0.029 (stat.)  & $\pm0.088$ (syst.) \tabularnewline
$\sigma\left(\nu\right)$ & 2.41  & $\pm$0.022 (stat.)  & $\pm0.231$ (syst.) \tabularnewline

$\Delta(\nu,\overline\nu)$   & 1.512  & $\pm$0.036 (stat.)  & $\pm$0.152 (syst.) \tabularnewline
$\Sigma(\nu,\overline\nu)$    & 3.311  & $\pm$0.036 (stat.)  & $\pm$0.318 (syst.) \tabularnewline
\hline 
\multicolumn{4}{c}{Ratios}\tabularnewline
\hline 
$R(\nu,\overline\nu)$  & 0.373  & $\pm$0.012 (stat.)  & $\pm$0.015 (syst.) \tabularnewline
$A(\nu,\overline\nu)$  & 0.457  & $\pm$0.012 (stat.)  & $\pm$0.17 (syst.) 
\tabularnewline
\hline
\hline 
\end{tabular}
\end{table*}

\begin{table*}
\caption{\label{tab:Full-phase-space_cross_sections}
Full phase space cross sections and ratio results extrapolated from restricted phase space measurements.}
\begin{tabular}{crcl}
\hline
\hline 
\multicolumn{3}{c}
{
Cross Sections {[}$\times10^{-39}$cm$^{2}$/nucleon{]} }\tabularnewline
\hline 
$\sigma\left(\overline{\nu}\right)$ & 1.71 & $\pm$0.29 (stat.+syst.)    \tabularnewline
$\sigma\left(\nu\right)$ & 7.07 & $\pm$1.20 (stat.+syst.)   \tabularnewline
\hline 
\multicolumn{3}{c}{Ratios}\tabularnewline
\hline 
$R(\nu,\overline\nu)$  & 0.242 & $\pm$0.058 (stat.+syst.)    \tabularnewline
\hline
\hline 
\end{tabular}

\end{table*}

\begin{figure*}
\includegraphics[scale=0.25]{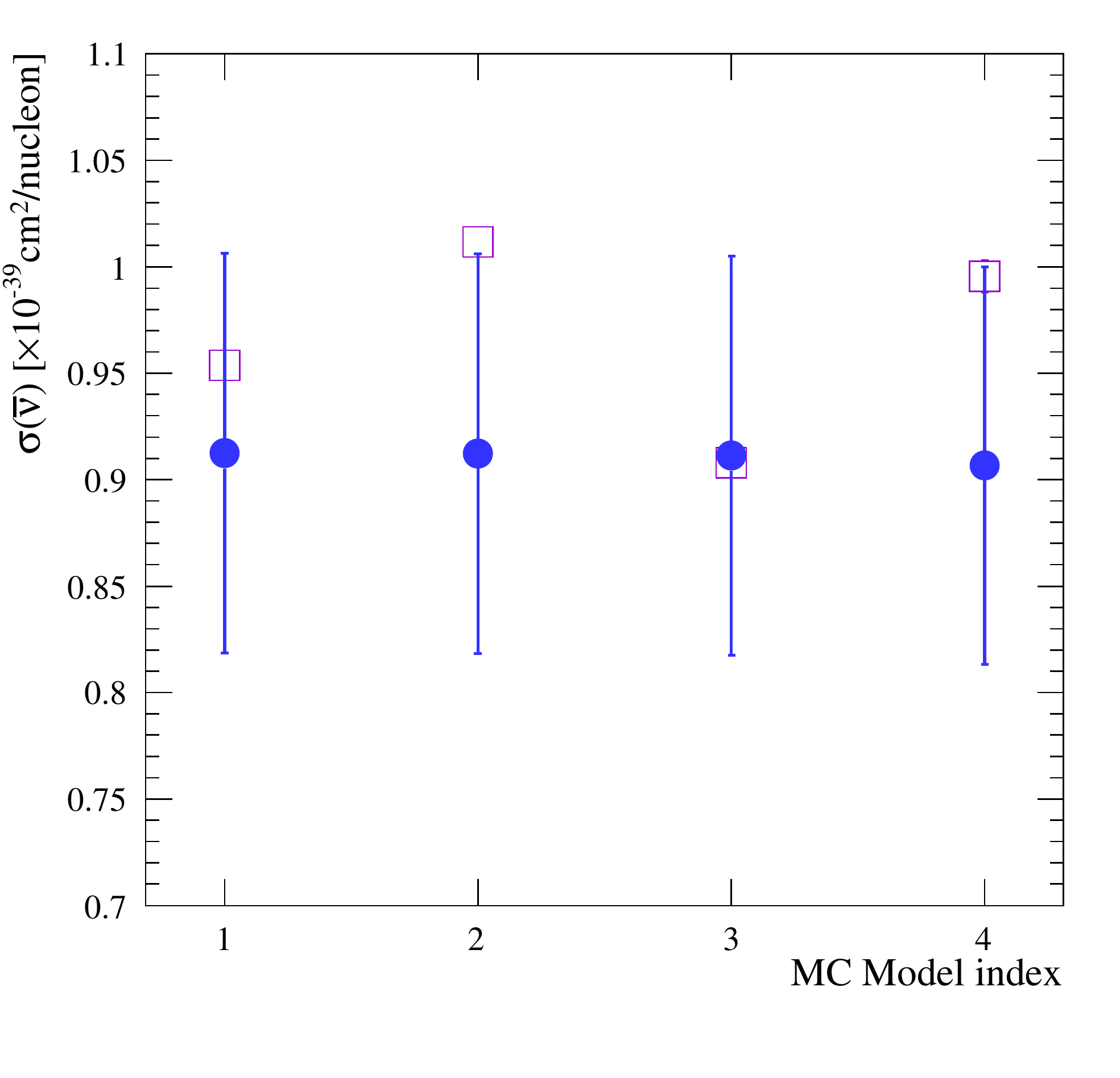}
\includegraphics[scale=0.25]{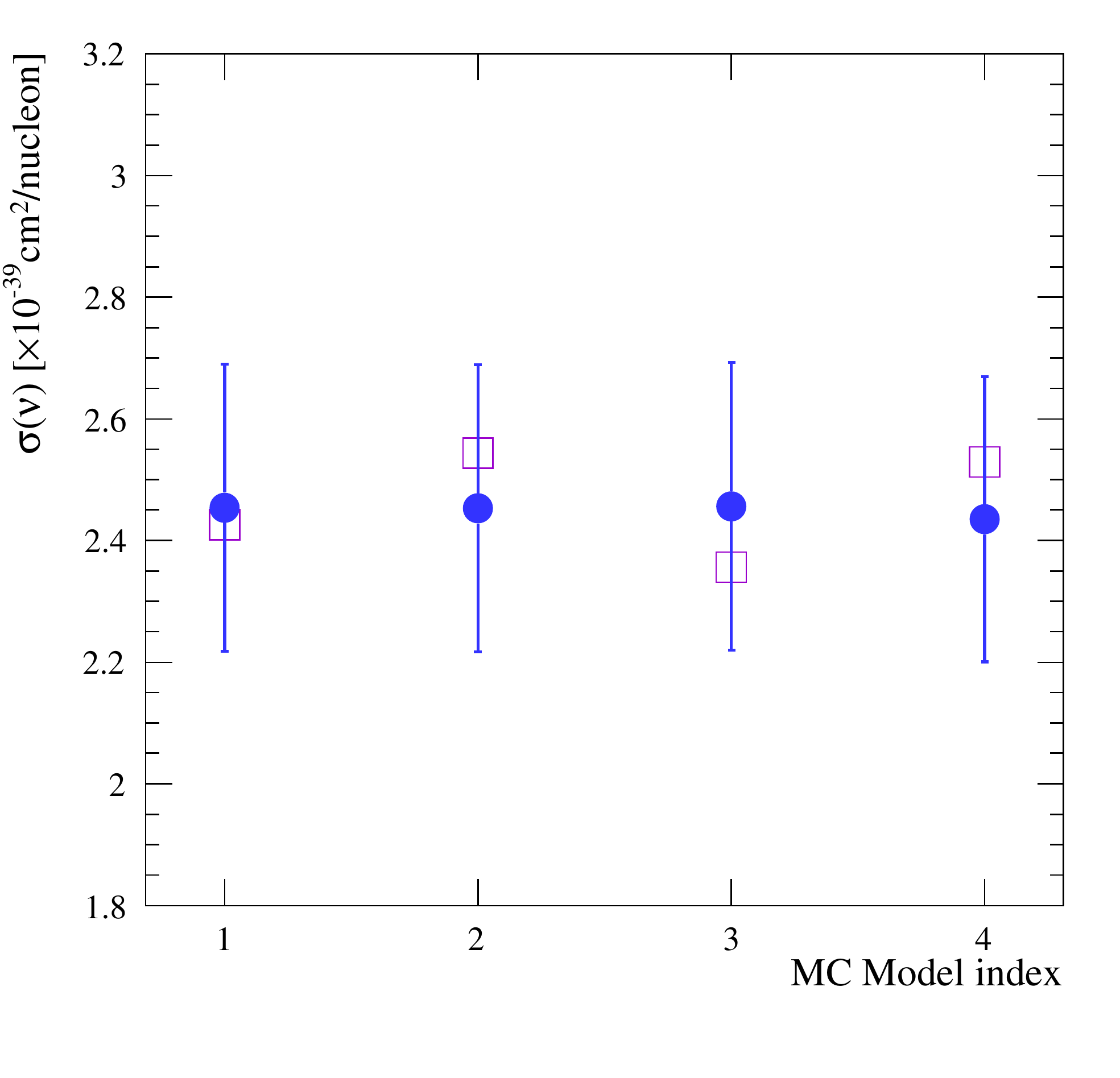}
\includegraphics[scale=0.25]{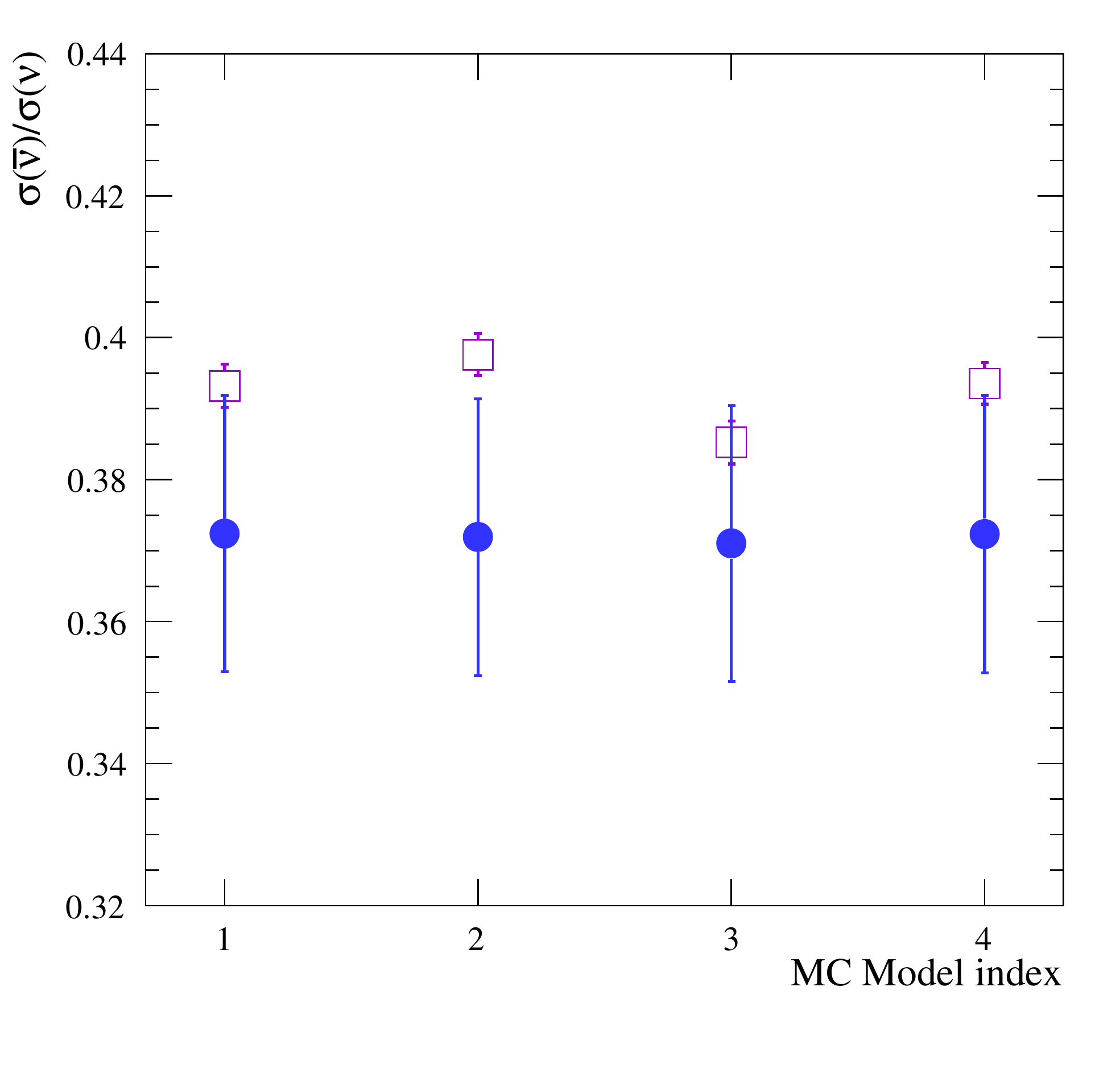}

\caption{\label{fig:2p2h-compare-1}  
Comparison of MC model 1-4 predictions, open squares with no errors bars, to data results, 
solid circles with error bars, in measurements 
of cross sections $\sigma(\bar\nu_\mu)$ [Left] 
and $\sigma(\nu_\mu)$ [Middle]  
and the $R$ ratio $\sigma(\bar\nu_\mu) / \sigma(\nu_\mu)$ [Right].
}
\end{figure*}

\begin{figure*}
\includegraphics[scale=0.25]{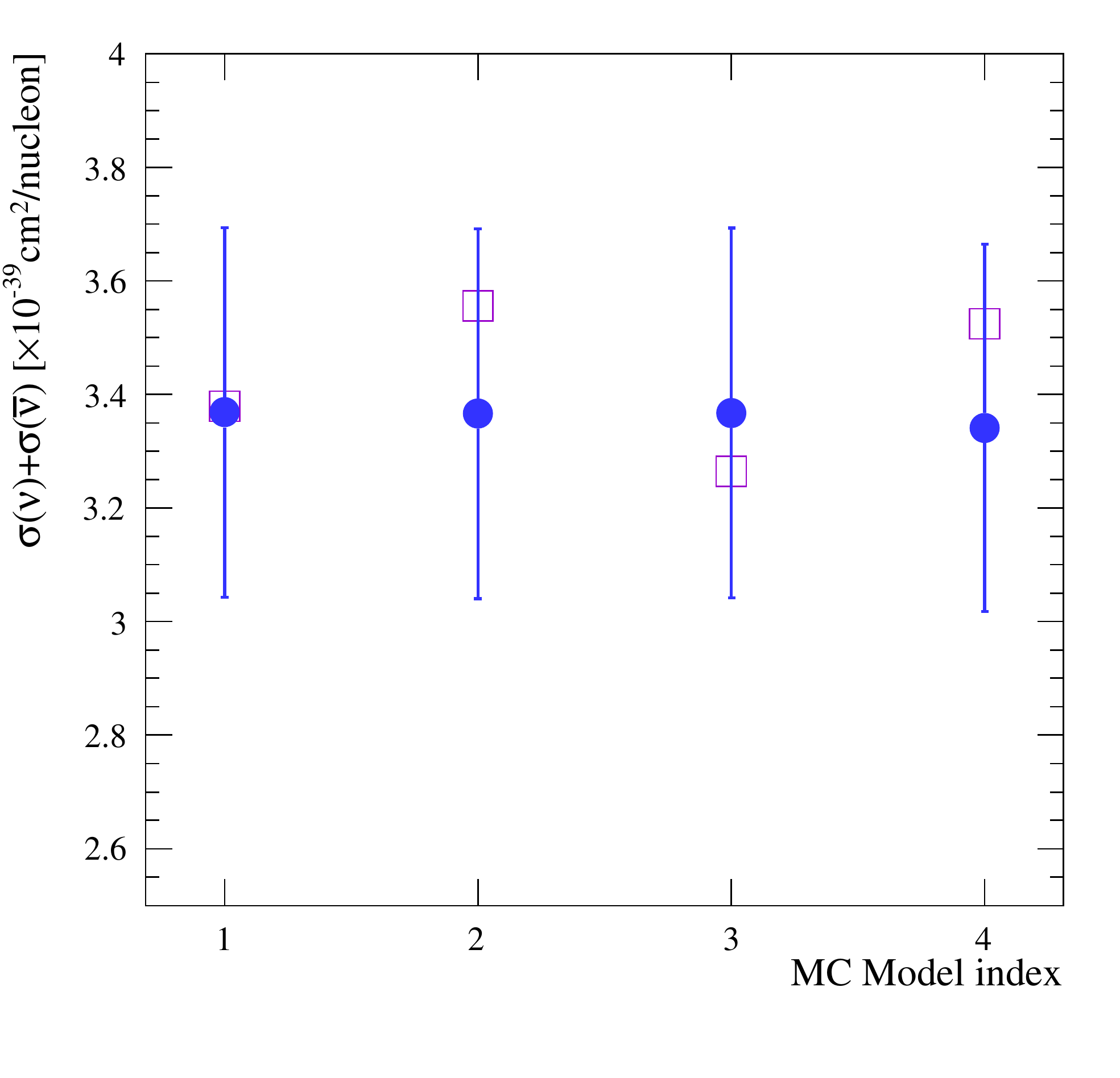}
\includegraphics[scale=0.25]{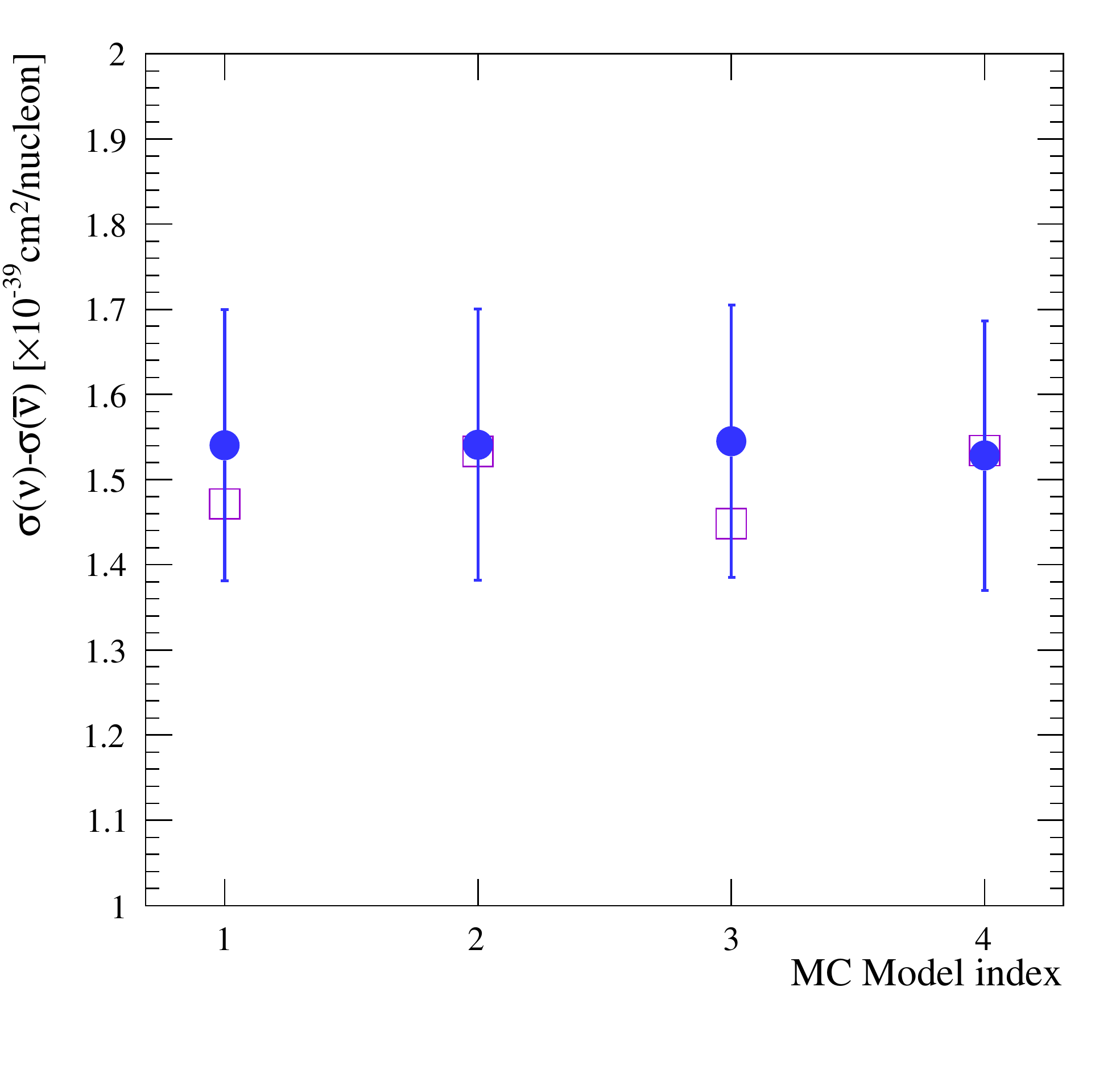}
\includegraphics[scale=0.25]{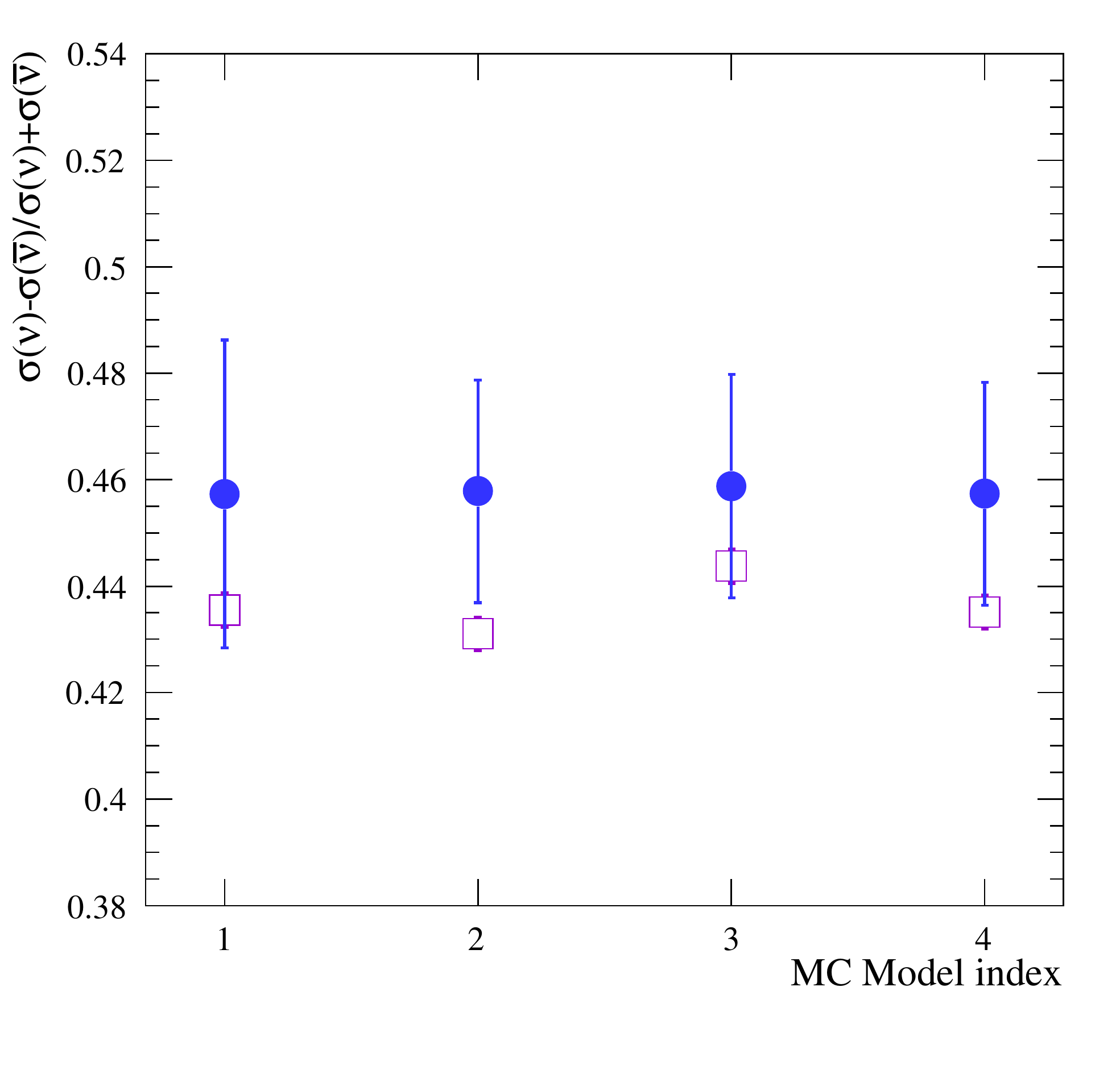}

\caption{\label{fig:2p2h-compare-2} 
 Comparison of MC model 1-4 predictions, open squares with no errors bars, to data results, 
solid circles with error bars, in measurements 
of the cross section sum $\Sigma=\sigma(\nu_\mu)+\sigma(\bar\nu_\mu)$ [Left], 
difference $\Delta=\sigma(\nu_\mu)-\sigma(\bar\nu_\mu)$ [Middle]  
and
asymmetry $A=( \sigma(\nu_\mu)-\sigma(\bar\nu_\mu) )$/$( \sigma(\nu_\mu)+\sigma(\bar\nu_\mu) )$ [Right].
}
\end{figure*}

\begin{table*}
\caption{
\label{tab:show-model3} 
The numerical values of model 3 predictions and 
the corresponding measurements shown in Figs. \ref{fig:2p2h-compare-1} 
 and \ref{fig:2p2h-compare-2}.
}

\begin{tabular}{lcccccc}
\hline
\hline
Model 3 &  $\sigma(\overline{\nu})$ & $\sigma(\nu)$ & $\Delta(\nu,\overline{\nu})$ & $\Sigma(\nu,\overline{\nu})$ & $R(\nu,\overline{\nu})$ & $A(\nu,\overline{\nu})$ \\
\hline
MC predictions   & $0.908$ & $2.36$ & $1.45$ & $3.26$ & $0.385$ & $0.444$ \\
Measurements & $0.911\pm0.094$ & $2.45\pm0.24$ & $1.55\pm0.16$ & $3.37\pm0.33$ & $0.371\pm0.019$ & $0.459\pm0.021$ \\

\hline
\hline
\end{tabular}
\end{table*}

\subsection{Discussion of results}

In this section we discuss how our results compare with NEUT predictions, previous measurements,
the impact on future CPV measurements and the multinucleon effects that can modify neutrino
cross sections.

We observe close agreement between the numbers of data events
and the NEUT MC generated events in both the unrestricted and restricted
phase space selected events. Using Table \ref{tab:input to final cross sections},
the data to MC ratios for the restricted phase space selection
for the FHC/RHC modes are 94.2\%/91.7\%.


We can compare our neutrino result to previous 
T2K publications that used the FGD sub-detector with a scintillator target.
The previous T2K flux averaged
CCINC \citep{t2k-carbon-ccinc} was 
 $(6.91\pm0.13{\rm (stat.)}\pm0.84{\rm (syst.)}) \times 10^{-39}$cm$^2$  per nucleon
and this is within systematic errors to our full phase space measurement
in Table \ref{tab:Full-phase-space_cross_sections}.
The published T2K CCQE \citep{t2k-cceq} and 
events of the charged current process that has no pions (CC0$\pi$) \citep{cc0pi}
flux averaged cross sections per nucleon are
$(3.83\pm0.55) \times 10^{-39}$cm$^2$ 
and
$(4.17\pm0.05\pm0.47) \times 10^{-39}$cm$^2$, respectively. 
In the context of the NEUT model, the CCINC results presented here are 
compatible with the CCQE and CC0pi results from these prior publications.
These full phase space neutrino results agree with the previous T2K measurements.

The near detector flux averaged uncertainties on the ratio of cross sections
and rates are useful to estimate the sensitivity of future CP conservation tests
in long baseline appearance experiments.  
The restricted phase space fractional systematic errors on  $R(\nu,\bar\nu)$  and $r(\nu,\bar\nu)$ 
are $4.0\%$ and $1.8\%$, respectively.  
These systematic errors on the near detector ratio measurements
are now due to many small errors less than $1\%$, so further substantial
improvements will be challenging.  
Although future measurements of appearance probabilities are likely to be
limited by statistical uncertainties on far detector
$\nu_e$ and $\bar\nu_e$ measurements,
the near detector uncertainties on
$\nu_\mu$ and $\bar\nu_\mu$ measurements 
may also limit the ultimate precision of future CPV tests.

The 2p2h models have been predicted \citep{martini-2} 
to affect the difference between the
\nuandnubar cross sections. The NEUT MC predictions
of the $\nu_\mu$ and \nubar cross sections, their difference and sum,
their ratio, and their asymmetry have been calculated in four models; 
(1) NEUT with a default Spectral Function \citep{sfmodel}, 
(2) RFG model, 
(3) RFG model with RPA corrections  and  
(4) RFG with RPA corrections and 2p2h interactions. 
The MC model (4) included
 2p2h effects in the NEUT MC generator from the model by 
Nieves \citep{nieves-1} and this model (4)
was also used to calculate the Table VI and VII results.
The six predicted (open squares) MC cross sections and their combinations of cross sections 
and 
the corresponding measurements (solid circles) for each model are displayed in 
Figs. \ref{fig:2p2h-compare-1} and \ref{fig:2p2h-compare-2}.
These models include additional nuclear effects such as 2p2h that make 
different predictions for neutrino and antineutrino enhancements to the cross section. 
We find different cross section combinations can help differentiate
the models and here we investigate a limited number of model combinations 
available in NEUT.
The measured cross sections are stable and have negligible changes with different models.
This demonstrates the efficiencies are similar in different models.
The observed \nubar cross section has slightly better agreement with model 3,
however the other models 1, 2 and 4 predictions are nearly all  
within 1 standard deviation of the data uncertainties. The numerical values of
the model 3 predictions and the data results are given in Table \ref{tab:show-model3}.
Although the uncertainty on our model combinations is relatively large, it is clear that 
with higher statistics, such comparisons will be valuable for model separation.

In future T2K measurements, more statistics, especially in the \nubar
mode, will enable differential water subtracted measurements in bins
of muon momentum and angle. After unfolding, the differential measurements
of ratios in particular, differences and sums are expected to provide improved estimates of
systematic uncertainties in future experimental CPV tests and better
tests of 2p2h models.

%% file: Sections/Conclusions.tex
\section{Conclusions}
In summary, the T2K experiment has measured charged current inclusive events, in a restricted phase
space of $\theta_{\mu}<32^{\circ}$ and $p_{\mu}>$500 MeV/c,
the flux averaged cross sections (cm$^{2}$ per nucleon) and ratio
of cross sections, as
\begin{equation}
\sigma(\overline{\nu})=\left(0.900\pm0.029{\rm (stat.)}\pm0.088{\rm (syst.)}\right)\times10^{-39},\label{eq:conclusion-1}
\end{equation}
\begin{equation}
\sigma(\nu)=\left(2.41\pm0.021{\rm (stat.)}\pm0.231{\rm (syst.)}\ \right)\times10^{-39}\label{eq:conclusion-2}
\end{equation}
and
\begin{equation}
R\left(\frac{\sigma(\overline{\nu})}{\sigma(\nu)}\right)=0.373\pm0.012{\rm (stat.)}\pm0.015{\rm (syst.).}\label{eq:conclusion-3}
\end{equation}
The \nubar inclusive cross section and the ratio $R$ results
are the first published measurements at \nuandnubar flux energies\citep{flux-model} below
1.5 GeV.  Although the current uncertainty on the different model combinations is relatively large, 
we expect future higher statistics comparisons will be valuable for model discrimination.